\documentclass[11pt]{article}
\pdfoutput=1

\usepackage{latexformat}
\usepackage{amsmath,bbm,array,amsfonts,wrapfig,lscape,float,mathtools,multirow,longtable,amsthm,upgreek,mathrsfs,graphicx}
\usepackage[dvipsnames]{xcolor}
\usepackage{stackrel}
\usepackage[all]{xy}
\usepackage[bottom]{footmisc}
\usepackage{mathrsfs}
\usepackage{tikz}
\usepackage{physics}
\usepackage{bm}
\usepackage[capitalize]{cleveref}
\usepackage{color}
\usepackage{enumerate}
\usetikzlibrary{decorations.pathreplacing}
\usepackage{diagbox}
\numberwithin{equation}{section}
\numberwithin{figure}{section}
\numberwithin{table}{section}
\usepackage{pgfplots}
\pgfplotsset{compat=1.14}
\usepackage{makecell}
\usepackage{lscape}
\usepackage[utf8]{inputenc}
\usepackage{mathtools}
\usepackage{slashed}
\usepackage{bigints}
\usepackage{comment}
\usepackage[toc, page]{appendix}

\usepackage{tocloft}

\allowdisplaybreaks

\graphicspath{{./figures/}}

\newtheorem{theorem}{Theorem}[section]

\renewcommand{\d}{\text{d}}
\newcommand{\D}{\text{D}}
\newcommand{\Pf}[0]{\text{Pf}}
\newcommand{\Pfaff}{\text{Pf}}

\title{Elliptic Genera of 2d $\mathcal{N}=(0,1)$ Gauge Theories}

	\author[a]{Jiakang Bao,}
        \author[a,b,c]{Masahito Yamazaki,}
        \author[a,b]{Dongao Zhou}
	
	\affiliation[a]{Graduate School of Physics,
        University of Tokyo, Tokyo 113-0033, Japan}
        \affiliation[b]{
		Kavli Institute for the Physics and Mathematics of the Universe,\\
		University of Tokyo, Kashiwa, Chiba 277-8583, Japan}
        \affiliation[c]{Trans-Scale Quantum Science Institute, 
        University of Tokyo, Tokyo 113-0033, Japan}
	
	\emailAdd{jiakang.bao@phys.s.u-tokyo.ac.jp}
        \emailAdd{masahito.yamazaki@ipmu.jp}
        \emailAdd{dongao.zhou@ipmu.jp}
	
	\preprint{
		\begin{flushright}
			
		\end{flushright}
	}
	
	\abstract{We derive an exact residue formula for the elliptic genera of 2d $\mathcal{N}=(0,1)$ gauge theories. We find a new residue prescription which recovers the Jeffery-Kirwan residue prescription for $\mathcal{N}=(0,2)$ theories. We apply the formula to the Gukov-Pei-Putrov model and analyze the phase structure of the theory.}

\begin{document}

\maketitle

\section{Introduction}\label{intro}
Supersymmetries have been one of the successful organizational principles in the study of quantum field theories. They are powerful tools to constrain the dynamics of the theories, to discuss non-perturbative dualities between theories, and even to solve the theories exactly.

While the analytic control from supersymmetry is more powerful for theories with more supersymmetries, one is also interested in reducing the number of supersymmetries. Theories with fewer supersymmetries have richer dynamics than their more supersymmetric counterparts, and we ultimately hope to learn general lessons about theories without any supersymmetries. For this reason, the case of 2d $\mathcal{N}=(0,1)$ theories is of particular interest, where we have the minimal supersymmetry consisting of one supercharge.
2d $\mathcal{N}=(0,1)$ theories are also of interest in the context of compactifications of the heterotic string theory \cite{Hull:1985jv, Hull:1986xn}.
More recently, $(0,1)$ theories have been studied in the context of the Stolz-Teichner conjecture and topological modular forms \cite{segal1987elliptic,stolz2004elliptic,stolz2011supersymmetric}.
Despite many-faceted interest in $(0,1)$ theories, many of the familiar arguments from supersymmetry (such as holomorphy) do not apply, and it is challenging to discuss their dynamics. It is therefore an important problem to extract any quantitative information for $(0,1)$ theories.

The goal of this paper is to provide such quantitative data for  $(0,1)$ theories. We shall derive an exact formula for the elliptic genus of $(0,1)$ gauged linear sigma models (GLSMs). In the literature, 
there are some discussions on the applications of the Atiyah-Bott localization \cite{Atiyah:1984px, berline1983zeros} for 2d $(0,1)$ theories \cite{Alvarez-Gaume:1983zxc, Alvarez-Gaume:1983sxa, Friedan:1983xr, Gukov:2019lzi}. Here, we shall compute the elliptic genera in the path-integral formulation. Our formula involves a novel residue prescription, which is reminiscent of, but crucially different from, the Jeffrey-Kirwan (JK) residue \cite{jeffrey1995localization} used for  $(0, 2)$ elliptic genera \cite{Benini:2013xpa}.

The rest of this paper is organized as follows.
In \cref{sec:setup}, we summarize the setups, notations, and assumptions.
In \cref{sec:ellgen}, we present our main results for the elliptic genera for $(0, 1)$ theories. The formula will be derived using supersymmetric localization of the path integral in \cref{sec:derivation}.
As a concrete application of our formula, we discuss in \cref{sec:GPP} the Gukov-Pei-Putrov (GPP) models. We conclude in \cref{sec:discussions} with summaries and open questions. We also include several appendices for technical materials.

\section{Setups and Notations}\label{sec:setup}
Let us first summarize the setups and notations for 2d $\mathcal{N}=(0,1)$ GLSMs. 
Some related details can be found, for example, in \cite{Sakamoto:1984zk,Siegel:1983es,Hull:1985jv,Brooks:1986uh,Brooks:1986gd,Gukov:2019lzi,Franco:2021ixh,Franco:2021vxq}.
While the contents are mostly standard, the discussion of the positive cones of the weight lattice in \cref{sec:reality} will be important in the next section. Moreover, in \cref{sec:assumptions}, we will state the important assumptions needed for the derivation in \cref{sec:derivation}.

\subsection{The \texorpdfstring{$(0, 1)$}{(0,1)} Supersymmetry}\label{(0,1)susy}
In $(0, 1)$ supersymmetry, we have three types of supermultiplets: scalar, Fermi, and vector multiplets. A real scalar multiplet is $\Phi= \phi + \theta^+\psi $, 
where $\phi$ is a scalar and $\psi$ is a right-moving Majorana-Weyl fermion. The supersymmetry transformation rules
under $\delta = \epsilon^+\mathcal{Q}_+$ are given by\footnote{We have introduced the lightcone coordinates $x^{\pm} = {(x^0\pm x^1)}/{\sqrt{2}}$  
such that the metric of the spacetime is $\d s^2 = -\d x^+\d x^-$.}
\begin{equation}
    \delta \phi = \epsilon^+ \psi \;,\quad\delta \psi  = \text{i}\,\epsilon^+ \partial_+ \phi\;.
\end{equation}
A Fermi multiplet is $\Gamma= \gamma+\theta^{+}D$,
where $\gamma$ is a left-moving Majorana-Weyl fermion and $D$ is an auxiliary scalar. The transformation rules of these fields are
\begin{equation}
    \delta \gamma = \epsilon^+D\;, \quad  \delta D= \text{i}\,\epsilon^{+} \partial_+ \gamma \;.
\end{equation}
A vector multiplet is given by $\Lambda_+=\theta^+A_+$, $\Lambda_-=A_-+\theta^+\lambda$ after gauge fixing, where $A$ is the gauge field and $\lambda$ is a left-moving Majorana-Weyl fermion under the adjoint representation of the gauge group. Consider the covariant derivatives $\nabla_+=\partial/\partial\theta^++\text{i}\,\theta^+\partial_++\text{i}\,\Lambda_+$ and $\nabla_-=\partial_-+\text{i}\,\Lambda_-$.
The field strength superfield 
is $\Sigma=\text{i}\,[\nabla_+,\nabla_-]=\lambda + \theta^{+} F_{+-}$, where $F_{+-} = \partial_+A_--\partial_-A_++[A_+,A_-]$. The variations of the components of $\Sigma$ are
\begin{equation}
    \delta\lambda_{-} = \epsilon^+ F_{+-}\;, \quad \delta F_{+-} = \text{i}\,\epsilon^+ \partial_{+} \lambda\;.
\end{equation}

\subsection{The Lagrangian}\label{sec:Lagrangian}
We shall consider an $\mathcal{N}=(0, 1)$ GLSM with gauge group $G$ and flavour symmetry group $G_F$. 
We denote the Cartan subgroup of $G$ and $G_F$ as $T$ and $T_F$, and their associated Lie algebras as $\mathfrak{t}$ and $\mathfrak{t}_F$.

In the theory, we have $(0, 1)$ vector multiplets associated with $G$, 
and in addition scalar multiplets (collectively denoted by $\Phi$) and Fermi multiplets (collectively denoted by $\Gamma$).
These multiplets are in real representations
$R^v, R^s, R^f$ of $G\times G_F$. We denote the associated weight spaces by $V^v, V^s, V^f$, whose dimensions we denote by $\textrm{dim}(G)$, 
$n_s$, and $n_f$.
Notice that $R^v$ is in a trivial representation of $G_F$, since the vector multiplet is not charged under the flavour symmetry.

To write down the Lagrangian, each weight space $V^{I=v, s, f}$ should be equipped with a non-degenerate bilinear pairing $\langle -, - \rangle_I$. The total Lagrangian density of the GLSM is given by
\begin{equation}
    \mathcal{L}_{\rm total} = \frac{1}{e^{2}} \mathcal{L}_{v}+ \frac{1}{g_s^2} \mathcal{L}_{s}+ \frac{1}{g_f^2} \mathcal{L}_{f}+ m\mathcal{L}_J \label{GLSM}
\end{equation}
with 
\begin{align}
&\mathcal{L}_{v} = \int \d \theta^+ \langle \Sigma, \nabla_{+} \Sigma \rangle_v  = \langle F_{+-}, F_{+-} \rangle_v + \text{i}\,\left\langle \lambda, \nabla_+^A\lambda \right\rangle_v\;, \\
&\mathcal{L}_{s} = \text{i}\,\int \d \theta^+  \langle \nabla_{+} \Phi, \nabla_{-} \Phi \rangle_s
=  -\left\langle \nabla_+^A\phi, \nabla_-^A\phi \right\rangle_s+ 
\text{i}\,\left\langle \psi , \nabla_-^A\psi  \right\rangle_s
+ \text{i}\,\langle \psi , \lambda\phi \rangle_s\;,\\
&\mathcal{L}_{f} = - \int \d \theta^+  \langle \Gamma, \nabla_{+} \Gamma \rangle_f =  \langle D, D \rangle_f+\text{i}\,\left\langle \gamma, \nabla_+^A\gamma\right\rangle_f\;,\\
&\mathcal{L}_J = \int \d \theta^+ \left\langle \Gamma, J(\Phi) \right\rangle_f  
= -\text{i}\,\left\langle \gamma ,\frac{\partial J(\phi)}{\partial\phi}\psi \right\rangle_f  +\langle D, J(\phi)\rangle_f\;,
\end{align}
where $\nabla_{\pm}^A=\partial_{\pm}+A_{\pm}$. The Lagrangians $\mathcal{L}_{v, s, f}$ are the Lagrangians for the vector, scalar, and Fermi multiplets, respectively. In addition, we have an extra term $\mathcal{L}_J$ representing the coupling between the scalar multiplet and the Fermi multiplet, 
where $J(\Phi)\in V_f$.

Integrating out the auxiliary field $D$ in \eqref{GLSM}, we obtain
\begin{equation}
	 -\frac{m^2 g_f^2}{4}\langle J(\phi), J(\phi) \rangle_f
     - \text{i}\,m \left\langle \gamma, \frac{\partial J(\phi)}{\partial \phi} \psi \right\rangle_f
     \;.
\end{equation}
This shows that the theory has $n_s$ bosons, $n_s$ right-moving fermions, and $n_f$ left-moving fermions. The interactions include a scalar potential and a Yukawa-type coupling.

The $J$-term can be expanded as
\begin{equation}
    J(\Phi) = J^{(0)}+J^{(1)}(\Phi) + J^{(2)} (\Phi,\Phi) + \dots\;,\label{eq:J_expand}
\end{equation}
where $J^{(k)}:(V^{s})^{\otimes k} \to V^{f}$ are multi-linear $G\times G_F$-equivariant maps.  
The $J^{(0)}$ term contributes a linear term $J^{(0)}D$ in the action
for each singlet $\Gamma = \gamma + \theta^+D$ in $V^{f}$. This provides a $(0,1)$ analogue of the FI term in the $(0,2)$ theory. For the $J^{(2)}$ term (with $J^{(2)}: (V^s)^{\otimes 2} \to V^f$), with the non-degenerate pairings defined on $V^{s}$ and $V^{f}$, it can be equivalently written as
a linear map
\begin{equation}
    \label{eq:J2_sharp}
    J^{(2) \#} : V^{f} \to V^{s}\otimes (V^{s})^* \cong \text{GL}(V^{s})
\end{equation}
such that $g_s^2\left\langle J^{(2)}(\Phi,\Phi),D\right\rangle_{f}=\left\langle\Phi, \text{i} J^{(2) \#}(D)\Phi\right\rangle_{s}$. 
The quadratic Lagrangian of the scalar multiplet can then be written as
\begin{equation}
    \mathcal{L}_{s} = \left\langle \phi, \left(\left(\overline{\partial}+u+\xi\right)\left(\partial + \overline{u}+\overline{\xi}\right) + \text{i} J^{(2) \#}(D) \right) \phi\right\rangle_s
    +\text{i} \left\langle\psi, \left(\partial + \overline{u}+\overline{\xi}\right)\psi \right\rangle_f \;,
\end{equation}
where we will hereafter set $m=1$, and we have denoted 
the elements of the Cartan subalgebras $\mathfrak{t}$ and $\mathfrak{t}_F$
by $u$ and $\xi$. These parameters represent background gauge fields in $T$ and $T_F$, and can be regarded as the chemical potentials for the $G$ and $G_F$ symmetries in the discussion of the elliptic genera.

\subsection{Reality and Weight Spaces}\label{sec:reality}
In a $(0, 1)$ theory, we have in general a real representation of $G\times G_F$. This is in contrast to 
the $(0,2)$ cases, where the matters are in complex representations due to the existence of the $\textup{U}(1)$ R-symmetry.

For a real representation, it is subtle to consider the weight space decomposition since $\mathbb{R}$ is not algebraically closed.
Given a real representation of a real Lie algebra (see \cref{realliealg} for details), we can think of its complexification as a representation of the corresponding complex Lie algebra: we take the complexified representations $R^{I=v, s, f}_{\mathbb{C}}$ with associated weight spaces $V^{I}_{\mathbb{C}}:= V^{I}\otimes_{\mathbb{R}}\mathbb{C}$. 
The weight space decomposition under the $G\times G_F$ symmetry can be expressed as
\begin{equation}\label{eq:weight_space}
  V^{I}_{\mathbb{C}}\cong\bigoplus_{i\in\mathfrak{P}^I}V^I_{i}\;.
\end{equation}
For each $i$, we have an associated weight $\rho_i=\left(\rho^G_i,\rho^F_i\right)$ consisting of $\rho^G_i \in \mathfrak{t}^{*} \simeq (\mathbb{C}^{*})^r$
(resp.~$\rho^F_i \in (\mathfrak{t}_F)^{*} \simeq (\mathbb{C}^{*})^{r_F}$)
for the maximal torus $T$ (resp.~$T_F$) of the $G$-action (resp.~$G_F$-action), so that  we have 
\begin{align}
    h\cdot v&=\rho^G_i(h)v\;, \quad h\in T,  \quad
    v\in V_i^I \;, \label{h_G}
    \\
    h\cdot v&=\rho^F_i(h)v\;, \quad h\in T_F, \quad
    v\in V_i^I \;, \label{h_F}
\end{align}
For a vector multiplet, we have $\rho^F_i=0$, and $\rho^G_i$ is a root of the Lie algebra of $G$.

Note that in \eqref{eq:weight_space} the weights $\rho_i$ are counted with multiplicities,
so that each $V^I_{i}$ is of complex dimension $1$
and the indices $i$ run over the set $\mathfrak{P}^I:=\{1,\dots,n_I\}$. 
(The $G\times G_F$ action in itself does not canonically specify such a decomposition
when the $\rho_i$'s have multiplicities, but we can choose any decomposition for our computational purposes.)

In addition to the action of $G\times G_F$,
we have an action of $J^{(2) \#}(D)$ \eqref{eq:J2_sharp} on $V^s$ for a given zero mode of the auxiliary field $D \in V^{f}_0$.
This action commutes with the maximal torus $T\times T_F$ of $G\times G_F$ since
\begin{equation}
    J^{(2) \#}(D) = J^{(2) \#}(hD) = hJ^{(2) \#}(D)h^{-1},\quad\forall\ h\in T\times T_F\;.
\end{equation}
The action of $J^{(2) \#}(D)$ can thus be simultaneously diagonalized with those of $T\times T_F$. Later in \cref{sec:assumptions}, we will state Assumption 3---this ensures that $J^{(2) \#}(D)$'s mutually commute for $D\in V^f_0$, where $V^f_0\subset V^f$ represents the Fermi multiplets uncharged (i.e.\ singlets) under $G\times G_F$. Then we can choose the decomposition
such that 
\begin{equation}
\label{D_J}
    J^{(2)\#}(D)\cdot v=Q_i(D)v\;, \quad D\in V^f_0 \;, \quad 
    v\in V_i^s\;, 
\end{equation}
where $Q_i\in\left(V^f_0\right)^*$.
Notice that there are similarities in 
\eqref{h_G}, \eqref{h_F}, and \eqref{D_J}, despite the fact that the $J$-term in itself does not represent a symmetry of the theory, at least in general.

Since we have complexified the representations, it is natural to return to real representations by choosing a real slice.
This is obvious in some cases --- for the action of $J^{(2) \#}(D)$, we find a reality condition $Q(D) \in \mathbb{R}$ since $J^{(2) \#}(D)$ is a real symmetric matrix.

We should note that we cannot always choose the weights
to take values in $\mathbb{R}$.
One can, however, choose a
``positive half of the weight lattice'' generically inside the weight lattice,
to pick up only positive weights.
Whenever the complex representation admits a real invariant structure, its weights always come in pairs\footnote{In particular, this is the case for the adjoint representation.}. The positive half picks up half of each pair.
Of course, this means that there are ambiguities in the choice of the positive half of the weight lattice,
to which we will come back when we discuss the elliptic genus.

When the complex representation does not admit an invariant real structure, the weights do not necessarily come in pairs. In such cases, the corresponding real representation has an invariant complex structure $\mathcal{J}$ (see \cref{theorem_real}). Then the real scalar or fermion, say $\phi_1$ or $\psi_1$, can be paired with $\phi_2=\mathcal{J}\phi_1$ or $\psi_2=\mathcal{J}\psi_1$ to form a complex field $\Phi$ or $\Psi$. Therefore, they could be combined into a complex representation, and we may take the \emph{full} weight space for the matter multiplet.

\subsection{Anomaly Cancellations}\label{anomalycancellations}
Since a $(0, 1)$ theory is in general chiral, we need to impose the anomaly cancellation conditions. The gauge and gauge-flavour  anomaly cancellation conditions are
\begin{align}
&\sum_{\textrm{fermions}}\gamma^3 J_G J_G = \sum_{\rho_G\in P^v} \rho_G(u)^2 
+ \sum_{\rho_G\in P^f} \rho_G(u)^2 
 - \sum_{\rho_G\in P^s} \rho_G(u)^2  = 0 \;,
 \label{eq:gauge_anomaly}\\
&\sum_{\textrm{fermions}}\gamma^3 J_G J_F=\sum_{(\rho_G, \rho_F)\in P^v} \rho_G(u) \rho_F(\xi) 
+ \sum_{(\rho_G, \rho_F)\in P^f} \rho_G(u) \rho_F(\xi) 
 - \sum_{(\rho_G, \rho_F) \in P^s} \rho_G(u) \rho_F(\xi)  = 0 \;,
\label{eq:grav_anomaly}
\end{align}
where $\gamma^3$ is the chirality matrix in two dimensions and $J_G$ (resp.~$J_F$) is the current for the gauge (resp.~flavour) symmetry $G$ (resp.~$G_F$). In this paper, we will not consider more subtle global anomalies. Note that we still allow non-trivial 't Hooft anomalies for $G_F$, as well as the gravitational anomalies.

\subsection{Technical Assumptions}\label{sec:assumptions}
For technical reasons to be explained in \cref{sec:derivation},
we assume in this paper that the $(0,1)$ GLSMs satisfy the following three properties. The first assumption concerns the $J$-term:
\begin{itemize}
    \item{\bf Assumption 1 (Genericity of Quadratic Superpotential)}
    We assume the $J$-term to be generic so that the coefficient matrix of the quadratic piece $J^{(2)}(\Phi,\Phi)$ in \eqref{eq:J_expand} is a full-rank matrix. 
    In other words, $J^{(2)}(\Phi,\Phi)$ is non-degenerate
    as a quadratic form.
\end{itemize}
This condition is natural since we are not aware of a non-renormalization theorem for the $(0,1)$ superpotential, and we expect this condition to be satisfied for a generic superpotential. 

The next assumption concerns the number of Fermi multiplets:
\begin{itemize}
    \item{\bf Assumption 2 (Zero Mode Cancellation)} The number of uncharged Fermi multiplets (under the maximal torus of $G\times G_F$) is the same as the rank of the gauge group. In other words, $\dim_{\mathbb{C}}\left(V^f_{0}\right) = \text{rank}(G)$. Moreover, there should be no such uncharged scalar multiplets.
\end{itemize}
We will show in our derivation that it is a \textit{sufficient} (but not necessary) condition for a non-trivial result. A more physical and geometrical argument will be presented in an upcoming paper. 
In the following we shall write\footnote{Notice that $V^f_0$ here stands for the subspace of weight 0, namely the space of fermions uncharged under the maximal torus of $G\times G_F$, and should not be confused with the notation $V_0$ for the real vector space in \cref{realliealg}.} 
\begin{align}
    b: =\dim_\mathbb{C}\left(V^f_0\right)\;, \quad
    r:=\text{rank}(G) \;.
\end{align}
In \cref{bneqr}, we will also mention the cases with $b\neq r$. In short, the elliptic genus would always vanish for $b<r$, while the analysis would be more complicated for $b>r$.

\begin{itemize}
    \item{\bf Assumption 3 (Simultaneous Diagonalizability)} The actions $J^{(2)\#}(D)$ for all $D\in V_0^f$ commute with each other for all $D\in V_0^f$.
\end{itemize}
In particular, this is the case when the weights in the weight space decomposition \eqref{eq:weight_space} 
have no multiplicities.

\begin{itemize}
\item{\bf Assumption 4 (Non-Degeneracy)} As we will discuss later in \cref{res}, the elliptic genus can be computed by evaluations of certain residues. We shall require the non-degeneracy condition for this residue: in the formula of the elliptic genus as a complex integral, each singularity surrounded by the contour is specified by an intersection of exactly $r$ hyperplanes from the denominator of the integrand.
\end{itemize}
We expect this assumption to be satisfied for a generic theory.

\section{Results for \texorpdfstring{$\mathcal{N}=(0,1)$}{N=(0,1)} Elliptic Genera}\label{sec:ellgen}
Let us first summarize our main result, i.e., the new exact expression for $\mathcal{N}=(0,1)$ elliptic genera.
The derivation via supersymmetric localization will be presented in \cref{sec:derivation}.

\subsection{Elliptic Genera}\label{ellipticgenera}

The elliptic genus is defined as
\begin{equation}
    I(\tau, \xi) := \text{Tr}_\mathcal{\text{RR}}\left[(-1)^F q^{H_L}\bar{q}^{H_R} \prod_{\alpha=1}^{r_F} \nu_{\alpha}^{K_{\alpha}} \right] \ \label{ellgendef},
\end{equation}
where the trace is over the Ramond-Ramond(RR)--sector of the Hilbert space.
We have $q=\text{e}^{2\pi i \tau}$ with $\tau$ being the modulus of the two-dimensional torus, and $K_{\alpha}$ are the generators of the Cartan subalgebra $T_F$ of the flavour symmetry group $G_F$ with the corresponding fugacities $\nu_{\alpha} = \text{e}^{2\pi i \xi_{\alpha}}$ ($\alpha=1, \dots, r_F)$. We shall often write below $\xi_{\alpha}$ collectively as $\xi$ (and likewise the gauge chemical potentials $u_{a=1, \dots, r}$ as $u$). Moreover, $H_L=(H+P)/2$ and $H_R=(H-P)/2$, where $H$ and $P$ are the Hamiltonian and the momentum. In the path-integral formalism, the elliptic genus describes the RR-sector path integral with flat background gauge fields coupled to the flavour currents.

The elliptic genus can be written as an integral of the 1-loop partition functions over the moduli space of flat gauge connections on the two-dimensional torus. 
This moduli space can be parametrized by the elements of the Cartan subgroup of the gauge group $G$ as $x_a = \text{e}^{2\pi\text{i}u_a}$, we have
\begin{equation}
    I(\tau, \xi)=\frac{1}{|W|}\sum_{u^*}\text{Res}_{u^*}(\eta)[Z_\text{1-loop}(\tau,u,\xi)]\;,
\end{equation}
where $W$ is the Weyl group of $G$.
In the following, we shall give the expression of $Z_\text{1-loop}$ and explain the precise specification of the residues.

\subsection{One-Loop Determinants}\label{oneloopdet}

The 1-loop determinant is a meromorphic $(r, 0)$-form and 
reads as a product $Z_\text{1-loop}=Z_vZ_fZ_s$, where $Z_v$ ($Z_f$, resp.~$Z_s$) collects the contributions from the vector (Fermi, resp.~scalar) multiplets:
\begin{align}
    &Z_v (\tau, u) = (-2\pi\text{i}\,\eta(\tau))^r\left(\prod_{i\in\mathfrak{P}^v_*}\frac{\text{i}\,\theta_1\left(\tau|\rho_i(u)\right)}{\eta(\tau)}\right)\left(\prod_{a=1}^r\d u_a\right)\;, \\
    &Z_{f} (\tau, u, \xi) = \prod_{i\in\mathfrak{P}^f_*}\frac{\text{i}\,\theta_1\left(\tau|\rho_i(u,\xi)\right)}{\eta(\tau)}\;, \\
    &Z_{s} (\tau, u, \xi) = \prod_{i\in\mathfrak{P}^s_*} \frac{\text{i}\,\eta(\tau)}{\theta_1\left(\tau|\rho_i(u,\xi)\right)}\;.
\end{align}
We will hereafter use the notation $\rho_i(u,\xi):= \rho^G_i(u) + \rho^F_i(\xi)$ for $\rho_i=\left(\rho^G_i, \rho^F_i\right)$, and $\mathfrak{P}^I_*$ indicates that the labels $i$ with $\rho_i=0$ are not included:
$\mathfrak{P}^I_* = \{i \in \{1, \dots, n\}  \, |\,  \rho_i\ne 0 \}$\footnote{For $i\in\mathfrak{P}^I$, $\rho_i(u)$ has no $\xi$ dependence since we have $\rho^F_i=0$ in this case as mentioned previously.}. Here, we have used the eta function and the theta function:
\begin{align}
    &\eta(\tau)=q^{1/24}\prod_{n=1}^\infty\left(1-q^n\right)\;,\\
    &\theta_1(\tau|z)=-\text{i}\,q^{1/8}y^{1/2}\prod_{n=1}^\infty\left(1-q^n\right)\left(1-yq^n\right)\left(1-y^{-1}q^{n-1}\right)\;,
\end{align}
with $y=\text{e}^{2\pi\text{i}z}$. 
For brevity, we shall also denote the theta function as $\theta_1(z)$ when this would not cause any confusion.

\subsection{Residues}\label{res}

To discuss the residue prescription, let us first recall the concept of the Grothendieck residue. (More details can be found, for example, in \cite{griffiths2014principles}.) Given a compact complex $r$-manifold $\mathfrak{M}$, consider a meromorphic $r$-form $\omega$ defined on some neighbourhood $U$ of $M$. Suppose that $\omega$ has divisors $\mathcal{D}_1,\,\dots,\,\mathcal{D}_r$ whose intersection $\mathcal{D}_1\cap\dots\cap\mathcal{D}_r$ is a discrete and finite set of points in $M$. 
In other words, given the open covering $\{U_i:=U-\mathcal{D}_i\}$ of $U-(\mathcal{D}_1\cap\dots\cap\mathcal{D}_r)$, the meromorphic $r$-form $\omega$ can be locally written as
\begin{equation}
    \omega=\frac{g(z)\d z_1\wedge\dots\wedge\d z_r}{f_1(z)\dots f_r(z)}
\end{equation}
with $f_1(z),\ \dots,\ f_r(z),\ g(z)$ holomorphic in $\overline{U}$. Moreover, $\mathcal{D}_i=\{z\in U\,|\,f_i(z)=0\}$. Then for a point $z^*$ in $\mathcal{D}_1\cap\dots\cap\mathcal{D}_r$, the Grothendieck residue is given by
\begin{equation}
    \text{Res}_{z^*}\,\omega
    :=\frac{1}{(2\pi\text{i})^r}\int_{\mathcal{C}_{z^*}}\omega\;,
\end{equation}
where $\mathcal{C}_{z^*}$ is the real $n$-cycle defined by $\mathcal{C}:=\{z\,|\,|f_i(z)|=\epsilon_i\}$ with sufficiently small $\epsilon_i$ surrounding $z^*$. The orientation of $\mathcal{C}$ is chosen by requiring $\d(\arg f_1)\wedge\dots\wedge\d(\arg f_r)\geq0$.

In our context, the compact manifold $\mathfrak{M}$ is (a cover of) the moduli space of flat connections:
\begin{equation}
    \mathfrak{M}\coloneq\mathfrak{t}_{\mathbb{C}}/\left(\text{Q}^\vee +\tau \text{Q}^\vee\right)\;,
    \label{eq:M_flat}
\end{equation}
where $\text{Q}^\vee$ is the coroot lattice. 
We consider the Grothendieck residue of $Z_\text{1-loop}$,
whose singularities determine the divisors 
\begin{align}
    \mathcal{D}_{\rho}:=\big\{\rho(u, \xi)=0 \big\} 
    \;, 
\end{align}
for $\rho \in P^s$.
The intersection of these divisors determines a set $\mathfrak{M}^*$ of isolated singularities $u^*$, where at least $r$ divisors meet.
Recall Assumption 4 regarding the non-degeneracy condition, i.e., the singularities are specified by an intersection of exactly $r$ divisors $u^* \in\mathcal{D}_1\cap\dots\cap\mathcal{D}_r$. In this case, let us choose the contour $\mathcal{C}$ surrounding $u^*$. We can then write the Grothendieck residue
\begin{align} 
     \text{Res}_{u^* }\, Z_\text{1-loop} = \frac{1}{(2\pi\text{i})^r} \int_{\mathcal{C}_{u^*}} Z_\text{1-loop}\;.
\end{align}

The definition of our residue requires more data.
Let us write the cone spanned by $\{Q_{i_1},\dots,Q_{i_r}\}$ as
\begin{equation}
    \text{Cone}(Q_{i_1}, \dots, Q_{i_r} )\coloneq\sum_{k=1}^r a_k Q_{i_k}\;,\quad a_k\geq0\;.
\end{equation}
We also consider all the cones generated by $r-1$ elements in $\{Q_{i_k}\}$, and denote the union of these cones as $\text{Cone}_\text{sing}(Q_{i_1}, \dots, Q_{i_r} )$. Then we
choose a covector $\eta\in(\mathbb{R}^r)^*$ such
that $\eta\notin\text{Cone}_\text{sing}(Q_{i_1}, \dots, Q_{i_r} )$. Note that different $u^*$ would have different $\text{Cone}(Q_{i_1}, \dots, Q_{i_r} )$ and $\text{Cone}_\text{sing}(Q_{i_1}, \dots, Q_{i_r} )$.

We can now define the residue $\text{Res}_{u^*}(\eta)$ to be
\begin{align}
    \text{Res}_{u^*}(\eta) [Z_\text{1-loop}] := \begin{cases}
     \displaystyle \frac{1}{(2\pi\text{i})^r} \int_\mathcal{C} Z_\text{1-loop}\;, &  \eta\in\text{Cone}(Q_{i_1}, \dots, Q_{i_r} ) \\ 
      0\;, & \eta\notin\text{Cone}(Q_{i_1}, \dots, Q_{i_r} )
      \end{cases}\;.
\end{align}

When one chooses a different half for the Pfaffian, the elliptic genus would only differ by a sign since the poles chosen based on $\eta$ would not be affected (which would not be the case if this were the JK residue). This is consistent with our discussions of the sign convention below. Although it is not evident from the mathematical expressions, due to the physical derivations in \cref{sec:derivation}, we expect that the elliptic genus would be independent of the choice of $\eta$.

\subsection{Comments}\label{comments}
We have seen the expression for computing the elliptic genera and discussed the ingredients in the formula. Let us now make some comments regarding the properties of the elliptic genera.

\paragraph{The overall sign} In our discussion above, we encountered an ambiguity in the choice of the positive cone of the weight space. Such an ambiguity leads to an overall sign ambiguity of the elliptic genus. In the path-integral derivation in \cref{sec:derivation}, we will see that this comes from the sign ambiguity of the Pfaffian, which can be regarded as a square root of the determinant. This sign ambiguity in defining the elliptic genus or the index was also pointed out in \cite{Gaiotto:2019gef,Witten:2019bou,Yonekura:2022reu}, and can arise from the generalized theta angle $\theta\in\text{Hom}\left(\Omega_2^\text{spin}(\text{pt}),\mathbb{R}/\mathbb{Z}\right)\cong\mathbb{Z}_2$ \cite{Kapustin:2014dxa}, where $\Omega_2^\text{spin}(\text{pt})$ is the bordism group of 2-manifolds $M$ with a spin structure and a sigma model map $f:M\rightarrow\text{pt}$. This sign 
is an additional piece of data beyond the Lagrangian that is required for the full specification of the theory.

\paragraph{Global gauge anomalies} 
There exists an extra requirement that the 1-loop determinant $Z_\text{1-loop}$ should be a well-defined function on the torus $T^2$. This is an incarnation of the absence of the global gauge anomaly under large gauge transformations. In particular, $Z_\text{1-loop}$ should be doubly periodic under the shifts of $u$, where the theta functions themselves transform
non-trivially as
\begin{equation}
    \theta_1(\tau|u+a+b\tau)=(-1)^{a+b}\text{e}^{-2\pi\text{i}\,bu-\pi\text{i}\,b^2\tau}\theta_1(\tau|u)
\end{equation}
for $a,b\in\mathbb{Z}$. This is, up to the overall sign ambiguity discussed above, ensured by the anomaly cancellation conditions in 
 \eqref{eq:gauge_anomaly}, \eqref{eq:grav_anomaly}.

\paragraph{Modular Transformation Properties} Under modular transformations, the elliptic genus transforms as
\begin{align}
I\left(\frac{a\tau + b}{c\tau +d}, \frac{\xi}{c\tau +d}\right)
=\epsilon(a, b,c, d)^{c_L-c_R}
\exp\left(-\frac{\pi\,\text{i}\,c}{ c\tau+d} (A_{FF})_{\alpha \beta} \xi^{\alpha} \xi^{\beta} \right) I(\tau, \xi)
\end{align}
for $\begin{pmatrix}a&b\\c&d\end{pmatrix}\in\text{SL}(2,\mathbb{Z})$. Here, $c_L-c_R$ is the gravitational anomaly, $A_{FF}$ are the flavour anomalies
\begin{align}
(A_{FF})_{\alpha \beta}: = \sum_{\rm fermions} \gamma^3 (J_F)_{\alpha} (J_F)_{\beta}\;,
\end{align}
and the multiplier system $\epsilon(a, b,c, d)$ represents an overall constant phase factor
defined by 
\begin{align}
\frac{\theta_1\left(\frac{a\tau + b}{c\tau +d}, \frac{\xi}{c\tau+d} \right)}{\eta\left(\frac{a\tau + b}{c\tau +d}\right)}
= \epsilon(a,b,c,d)\, 
\exp\left(-\frac{\pi\,\text{i}\,  c }{ c\tau+d} \xi^2 \right)
\frac{\theta_1(\tau, \xi)}{\eta(\tau)}\;.
\end{align}

\paragraph{Global residue theorem} The global residue theorem states that the sum of the Grothendieck residues over all the singular points is zero:
\begin{equation}
    \sum_{u^* \in \mathfrak{M}^*}\text{Res}_{u^*}\, Z_\text{1-loop} =0\;.
\end{equation}
Suppose that there is a phase of the theory where all $u^*\in \mathfrak{M}^*$ 
sit inside the positive cone. Then the residue theorem implies that the elliptic genus is zero in this case, and it could be possible that such a phase can be continuously deformed to the phase where supersymmetry is broken\footnote{In general, the vanishing of the elliptic genus is a sufficient but not a necessary condition for the breaking of the supersymmetry. Nevertheless, for the examples considered in this paper, we shall provide some further evidence (analogous to the $(0,2)$ arguments in \cite[Section 5.4.4]{Melnikov:2019tpl}) in \cref{sec:GPP} and \cref{dynamics}.}. We will encounter an example later in \cref{sec:GPP}.

\subsection{Comparisons with  \texorpdfstring{$(0,2)$}{(0,2)} Theories}\label{comparisons}

Our formula includes and generalizes the previous results for the $(0, 2)$ theories,
where $(0,2)$ multiplets decompose into $(0,1)$ multiplets as 
\begin{equation}
    \begin{tabular}{ccc}
        $(0,2)$ & $\rightarrow$ & $(0,1)$ \\ \hline
        vector & $\rightarrow$ & vector $+$ Fermi \\
        Fermi & $\rightarrow$ & Fermi $+$ Fermi \\
        chiral & $\rightarrow$ & scalar $+$ scalar
    \end{tabular}
\end{equation}
so that the one-loop determinant for a $(0, 2)$ multiplet is a product of the one-determinants for two $(0, 1)$ multiplets.
We can verify that the product of the 1-loop determinants of the two $(0,1)$ multiplets indeed recovers the 1-loop determinant of the corresponding $(0,2)$ multiplets in \cite{Benini:2013xpa}.

Let us emphasize that the residue we defined above is in general {\it not} the JK residue,
and requires a {\it new} prescription for the residue.
The standard JK residue in our language corresponds to the 
special case where $Q_{i}$ coincides with $\rho^G_{i}$. For a general $(0, 1)$ theory, we have $Q_i\neq\rho^G_i$ while for $(0, 2)$ theories we have $Q_i=\rho^G_i$, reproducing the JK residue formula of \cite{Benini:2013xpa}.

In the $(0,2)$ case, there is a $(0,1)$ Fermi multiplet in the adjoint representation of the gauge group with $b=r$ zero modes, and the quadratic $J$-terms are the standard Yukawa couplings between the gaugini and the scalar multiplets given by the moment map of the $G$-action. Then the elliptic genus would be given by the standard JK residue, where $Q_i$ are given by the weights $\rho_G^i$ of the matter representations of $G$. By contrast, a $(0,1)$ vector multiplet does not have any auxiliary fields, and the zero mode pairings require the auxiliary fields from the Fermi multiplets (and hence the Assumption 2 introduced previously). This means the poles that should be taken would now depend on the choice of the $(0, 1)$ superpotential \eqref{eq:J_expand} in the $J$-term, and $Q_i$ are no longer the charges of the scalar fields under the action of the gauge group $G$: they are extra data from the quadratic $J$-terms that couple the Fermi multiplet zero modes to the bosonic multiplets.
 
\section{The Path Integral Localization}\label{sec:derivation}
In this section, we derive the formula for the $\mathcal{N}=(0,1)$ elliptic genera in the path-integral formalism via supersymmetric localization. Concretely, we shall compute the flavoured/equivariant elliptic genus of the GLSM defined in \cref{sec:Lagrangian}. After Wick rotation\footnote{We remark that there are no Majorana-Weyl spinors in the 2d Euclidean space. However, we may treat the Wick rotation as a formal operation preserving the path integral. In particular, this will not affect the definition of the fermion Pfaffian. To see this, consider the action of a Majorana fermion given by $S_0=\int_M\overline{\psi}\,\text{i}\,\slashed{\D}\,\psi$. After the Wick rotation with $\text{i}\,\slashed{\D}\rightarrow\slashed{\D}_E$, we have $S_E=\int_M\psi(C\slashed{\D}_E)\psi$, where $C$ is the charge conjugation matrix that is invariant under Wick rotation. As the conjugate of $\psi$ does not appear in the expression, the path integral would still formally give $\int\D\psi\,\text{e}^{-S_E}=\Pf(C\slashed{\D}_E)$. Notice that we are not distinguishing $\slashed{\D}$ and $\slashed{\D}_E$ in the main context.}, the Hamiltonian formula \eqref{ellgendef} can be recast in the path integral formalism:
\begin{equation}
        I(\tau,\xi)=\int_{\text{PBC}}[\d c]\,[\d\overline{c}]\,[\d\phi]\,[\d\psi]\,[\d\gamma]\,[\d D]\,[\d A]\,\left[\d \overline{A}\right]\,[\d \lambda]\,\exp(-S_\text{GLSM}-S_\text{gf})\;,
\end{equation}
where $S_\text{gf}$ is the gauge fixing term and PBC stands for the periodic boundary condition.

\subsection{The Path Integral Formalism}\label{pathintegral}
We can argue for the invariance of $I(\tau,\xi)$ under the reparameterization of parameters by the standard supersymmetry localization technique. In fact, we have
\begin{equation}
    S_\text{GLSM}=\int\d\theta^+\,\ell\;,
\end{equation}
which holds for any superfield Lagrangian in $S_\text{GLSM}$. (We can write $\text{D}_+$ instead of the gauge-covariant derivative $\mathcal{D}_{+}$ since $\ell$ is not charged under the gauge symmetry.) Therefore, a variation with respect to any parameter in it gives
\begin{equation}
    \delta I(\tau,\xi)\propto\int[\dots]\,\delta _{\mathcal{Q}_+}i^*\ell \exp(-S_\text{GLSM}-S_\text{gf})=\int[\dots]\,\delta _{\mathcal{Q}_+}(i^*\ell \exp(-S_\text{GLSM}-S_\text{gf}))=0\;.
\end{equation}
This holds as long as the measure $[\dots]$ is non-anomalous invariant under supersymmetry and no extra contributions originate from the non-compact directions in the field configuration space\footnote{As we will see later, one of such possibilities would be the bosonic zero modes appearing at specific loci in the moduli space.}.

In the weak coupling limit $e,\ g_s,\ g_f\to 0$, the path integral localizes to the BPS configurations
\begin{equation}
   (\partial+A)\phi=0\;,\quad F_{+-}=0\;.
\end{equation}
The solutions to the first equation are isolated at generic points, but may develop zero modes for specific points of the moduli space. The second equation gives the moduli space of flat connections on the worldsheet torus, which can be identified as $\mathfrak{M}/W$ with $\mathfrak{M}$ defined in \eqref{eq:M_flat}. We parametrize the cover $\mathfrak{M}$ by
\begin{equation}
    u = u_aT^a = \oint_{\mathsf{a}} A - \tau\oint_{\mathsf{b}} A \;,
\end{equation}
where $T^a$ are generators of the Cartan subalgebra and $\mathsf{a}, \mathsf{b}$ are the standard basis of non-contractible cycles on $T^2$. Similarly, we parametrize the flat connection of flavour symmetry by
\begin{equation}
    \xi = \xi_{\alpha} T_F^{\alpha} = \oint_{\mathsf{a}} A_F - \tau \oint_{\mathsf{b}} A_F \;.
\end{equation}
These parameters are to be identified with the chemical potentials for the gauge/flavour in the Hamiltonian formulation, denoted previously by the same symbols.

The Fermi multiplets may also develop zero modes that can be soaked up by the Yukawa couplings in the superpotential. These zero modes will be important in the following analysis. We shall divide them into \textit{generic zero modes} and \textit{coincident zero modes}. The generic zero modes are those existing in the whole moduli space of the gauge fields, while the coincident zero modes are those only appearing in some special subspaces of the moduli space. Suppose that a field has weight $\rho$ under $G\times G_F$. Then the condition for the existence of a zero-mode solution is
\begin{equation}
    \rho(u,\xi)\equiv0\quad\text{mod}\quad2\pi\text{i}(\mathbb{Z}+\tau\mathbb{Z})\;.
\end{equation}
Therefore, a generic zero mode exists only when $\rho(u, \xi)=0$, i.e., when the field is charged under neither the gauge nor the flavour symmetry.

To perform the path integral in the weak coupling limit, we shall separate the fields into zero and non-zero modes:
\begin{align}
    \begin{split}
        &A = u + eA'\;,\quad \bar{A} = \overline{u} + e\overline{A}'\;,\quad \lambda = \lambda_0 + e\lambda'\;, \\
        &\phi = e\phi'\;,\quad \psi = e\psi'\;,\quad \gamma = \gamma_0+e\gamma'\;,\quad D = D_0 + eD'\;.
    \end{split}
\end{align}
We have assumed that there are no generic scalar zero modes since they would break the compactness of the theory and make the elliptic genus ill-defined (without suitable regularizations).

\paragraph{Vector multiplets} The 1-loop contribution from $A'$ will be cancelled by the ghost fields, since there are no local degrees of freedom in two-dimensional gauge fields. With the periodic boundary condition in both directions, it will have $r = \text{rank}(G)$ generic zero modes. The 1-loop determinant is
\begin{equation}
    Z_v = \text{Pf}^{~'}_{V_\text{}\otimes C^\infty(T^2)}\left(\overline{\partial}+u\right) = (-2\pi\text{i}\,\eta(\tau))^r\left(\prod_{i\in\mathfrak{P}^v_*}\frac{\text{i}\,\theta_1(\tau|\rho_i(u))}{\eta(\tau)}\right)\left(\prod_{a=1}^r\d u_a\right)\;,
\end{equation}
where the primed Pfaffian means that we have excluded the zero-mode contributions.

\paragraph{Fermi multiplets} The Fermi multiplet is coupled to both the flavour and the gauge symmetries. By our Assumption 2, $\dim_\mathbb{C}\left(V^f_0\right)=r$. Since a Fermi multiplet has an additional auxiliary field $D$, there are also $r$ zero modes $D_0$ from the auxiliary fields. After excluding the zero modes, the 1-loop determinant is
\begin{equation}
    Z_{f} = \text{Pf}^{~'}_{V^{s}\otimes C^\infty(T^2)} \left(\overline{\partial}+u+\xi\right) = \prod_{i\in\mathfrak{P}^f_*}\frac{\text{i}\,\theta_1(\tau|\rho_i(u,\xi))}{\eta(\tau)}\;.
\end{equation}

\paragraph{Scalar multiplets} For the scalar multiplets, a new feature of the 1-loop determinant is the contribution of the $D\phi\phi$ terms coming from the superpotential coupling. Let us expand the superpotential $J$ as in \eqref{eq:J_expand}.
For a generic superpotential, it is not hard to see that higher $J^{(k)}$ terms would get suppressed by $e^k$ or $e^{k+1}$, depending on whether we are considering the $D\phi^k$ or the $D'\phi^k$ coupling. Therefore, we shall only keep the terms up to $e^2$ and further eliminate the $J^{(1)}$ term by a constant shift\footnote{In order to keep the $G\times G_F$ action on $V^{n_s}$ linear under the shift, we need to expand the $J$-term at $\phi_c\in\left(V^{n_s}\right)^{T\times T_F}$ fixed under the action of the Cartan subgroup $T\times T_F$. However, by Assumption 2 there would not be any generic zero modes of scalar multiplets. Hence, combined with the linearity, there is always one and only one fixed point, which is the origin $\phi_c = 0$.}.

The 1-loop determinant is
\begin{align}
    Z_{s} 
    &= \frac{\text{Pf}_{V^{s}\otimes C^\infty(T^2)} \left(\partial + \overline{u}+\overline{\xi}\right)}{\sqrt{\text{Det}_{V^{s}\otimes C^\infty(T^2) }\left(\left(\overline{\partial}+u+\xi\right)\left(\partial + \overline{u}+\overline{\xi}\right) + \text{i}\, J^{(2) \#}(D)\right)}}
    \nonumber \\
    &= \prod_{i\in\mathfrak{P}^s_*} \prod_{m,n\in\mathbb{Z}} \frac{n+m\overline{\tau}+\rho_i\left(\overline{u},\overline{\xi}\right)}{|n+m\tau +\rho_i(u,\xi)|^2+\text{i}\,Q_i(D)}\;.\label{Jprimedeigen}
\end{align}
When $D=0$, this simplifies to
\begin{equation}
    Z_{s} = \prod_{i\in\mathfrak{P}^s_*} \frac{\text{i}\,\eta(\tau)}{\theta_1(\tau| \rho_i(u,\xi))}\;.\label{Scalar_Det}
\end{equation}

We emphasize that the $J^{(2) \#}(D)$ term provides an effective mass term thanks to Assumption 1. This regulates the divergence of the determinant when $n+m\tau + \rho_i(u,\xi)=0$, where the theory develops coincident bosonic zero modes, causing non-compactness and making the elliptic genus ill-defined. Since we shall finally integrate over the whole moduli space, such ill-defined points are unavoidable. Therefore, generic Fermi multiplet zero modes and quadratic superpotential couplings to the scalar multiplets are necessary in our computations.

\paragraph{Interaction terms} The existence of the generic fermionic zero modes would make the naive 1-loop partition function vanish identically. However, there are Yukawa couplings in our action coming from the gauge-matter couplings and the superpotential, which will soak up the zero modes and lead to non-trivial results. These terms can be written as
\begin{equation}
    S_\text{int} = \int \d^2x \  \langle\psi,\lambda \cdot\phi \rangle_f + \left\langle \gamma, \frac{\partial J^{(2)}}{\partial \phi}\psi \right\rangle_{f}\;.
\end{equation}
Gathering all the terms, we have\footnote{Henceforth, we shall suppress the subscripts 0 and use $\gamma,\ \lambda,\ \dots$ themselves to represent the corresponding zero modes.}
\begin{align}
    I(\tau,\xi) =& \int \d^ru\ \d^r\overline{u}\ \d^r\lambda\ \d^r\gamma\ \d^rD \nonumber \\ 
    &\left\langle\exp(-S_\text{int}) \right\rangle_0 Z_s\left(u,\overline{u},D\right) Z_f(u)Z_v(u)\exp\left(- \int \d^2z \ \left(\frac{1}{2e^2}D^2 - \text{i}\,J^{(0)}(D) \right) \right)\;.\label{Ell-middle}
\end{align}
Here, $\langle\dots\rangle_0$ denotes the vacuum expectation value. We can compute this using the standard perturbative techniques. Then the fermionic zero mode integral will extract the terms with $r$ external legs of $\lambda$ and $\gamma$.

\subsection{The Integration Region}\label{region}
As was pointed out in the discussions of \eqref{Scalar_Det}, the 1-loop determinant of the bosonic fields in the scalar multiplets may diverge at specific loci in the moduli space of flat connections. We shall denote these loci by $H_i$:
\begin{equation}
    H_i \coloneq \{u_*\in \mathfrak{M} \ |\ \exists\ i\in\mathfrak{P}^s_* \text{ s.t. }\exp(\text{i}\rho_i(u_*,\xi))=1 \}\;.
\end{equation}
At these points, there exist bosonic zero modes satisfying the periodic boundary condition. As a result, the 1-loop determinant would diverge if we take $D=0$. However, the auxiliary field zero modes $D$ would regularize this divergence. To see this, we introduce the following tubular neighborhoods around $H_i$:
\begin{equation}
    H_i^\epsilon \coloneq \{u_*\in \mathfrak{M} \ | \ \exists\ i\in\mathfrak{P}^s_* \text{ s.t. } |\exp(\text{i}\rho_i(u_*,\xi))| < \exp(\epsilon) \}\;.
\end{equation}
Suppose that there are $k$ bosonic zero modes in the region $H_{{j}}^\epsilon$. Integrating out $D$ gives a potential $J(\phi_0)^2$, and we estimate
\begin{equation}
    \int_{H_i^\epsilon}\d^2u \int \d^k\phi_0\ \exp\left(-\frac{1}{g_s^2}|\rho_i(u-u_*)|^2 |\phi_0|^2 -  g_f^2J(\phi_0)^2 \right) \sim C (g_s)^k(g_f)^{-k}\epsilon^2\;,
\end{equation}
where $C$ is some constant that does not depend on $g_s$, $g_f$ or $\epsilon$. Note that we only consider the bosonic fields in the scalar multiplets, and the contributions from the corresponding fermionic fields would cancel the $g_s$ factor in this expression. Therefore, we conclude that the integration would be safe if we take the limit $\epsilon\to0$ appropriately such that $\epsilon^2(g_f)^{-k} \to 0$.

Denoting the union of $H_{{j}}^\epsilon$ as $H^\epsilon$,  we can separate the integration over $\d^{2r}u$ into two parts:
\begin{equation}
    \int_{\mathfrak{M}} =  \int_{\mathfrak{M}\backslash H^\epsilon} + \int _{H^\epsilon}\;.
\end{equation}
By taking the limit as above, we find
\begin{equation}
    \int_{\mathfrak{M}}\dots  =  \lim_{\epsilon\to0} \int_{\mathfrak{M  }\backslash H^\epsilon}\dots\;,
\end{equation}
to verify the absence of the contributions from the singular loci $H^{\epsilon}$ in the limit.
The remaining problem is to evaluate the 
integral in the remaining region ${\mathfrak{M  }\backslash H^\epsilon}$.

\subsection{The Stokes Relation and the \texorpdfstring{$D$-Contour}{D-Contour}}\label{StokesDcontour}
Computing the vacuum expectation value in \eqref{Ell-middle} is not a simple task since it depends on the details of the $J^{(2)}$ term in the superpotential. Fortunately, there is an argument (See, for example, \cite{Benini:2015noa}) which can help us simplify it.

Let us start with the $(0,1)$ supersymmetry transformation rules of the zero modes:
\begin{equation}
    \delta_{\mathcal{Q}_+} u = \lambda\;,\quad\delta_{\mathcal{Q}_+}\lambda = 0\;,\quad\delta_{\mathcal{Q}_+}\bar{u} =0\;,\quad\delta_{\mathcal{Q}_+} \gamma = -D\;,\quad\delta_{\mathcal{Q}_+} D = 0 \;.
\end{equation}
We have used the fact that the zero modes are constants on $T^2$. The zero-mode measure is invariant since there are no diagonal terms. Therefore, the whole integrand in \eqref{Ell-middle} should also be invariant under this transformation. This immediately gives the following relation:
\begin{equation}
    \left(D^a\frac{\partial}{\partial \gamma^a} - \lambda^a \frac{\partial}{\partial u^a} \right) Z(u,\bar{u},\lambda,\gamma,D)=0\;.\label{Stokes}
\end{equation}

In parallel with the proposals in \cite{Benini:2013nda,Benini:2013xpa}, given a set of charge vectors $\{\mathsf{Q}_1,\dots,\mathsf{Q}_s\}$ \emph{dictated by the quadratic $J$-terms}, we construct an $(r,r-s,r)$-form in the $(u,\bar{u},D)$-space (recall that $b=r$ based on Assumption 2):
\begin{equation}
    \mu_{\mathsf{Q}_1,\dots,\mathsf{Q}_s} \coloneq \d^ru \wedge \Omega_{a_1\dots a_s}\wedge \frac{\mathsf{Q}_1^{a_1}\dots \mathsf{Q}_{s}^{a_s}}{\mathsf{Q}_1(D)\dots \mathsf{Q}_s(D)} \d^b D\;.\label{mu-Q-def}
\end{equation}
Here,
\begin{equation}
\Omega_{a_1 \ldots a_s} \coloneq \left.\frac{1}{(r-s)!^2} \d \overline{u}^{c_1} \wedge \dots \wedge \d \overline{u}^{c_{r-s}}\ \epsilon_{b_1 \ldots b_{r-s} a_1 \ldots a_s} \frac{\partial^{2(r-s)}}{\partial \lambda^{c_1} \partial \gamma^{b_1} \ldots \partial \lambda^{c_{r-s}} \partial \gamma^{b_{r-s}}} Z\ \right|_{\lambda=\gamma=0}\;.
\end{equation}

We follow the arguments similar to those in \cite{Benini:2013xpa}. First, \eqref{Stokes} implies that
\begin{equation}
    \overline{\partial}\Omega_{a_1\dots a_s} = (-1)^{r-s}\sum_{i=1}^s (-1)^{i-1} D_{[a_i}\Omega_{a_1\dots\widehat{a}_i\dots a_n]}\;,\label{partial-D-relation}
\end{equation}
where the hat indicates the omission of the corresponding subscript. It then follows that
\begin{equation}
    \d\mu_{\mathsf{Q}_0,\dots,\mathsf{Q}_s} = \sum_{i=0}^s (-1)^{s-i}\mu_{\mathsf{Q}_0,\dots,\widehat{\mathsf{Q}}_i,\dots,
    \mathsf{Q}_s} \ \label{iteration_formula}
\end{equation}
In the situations with zero or $r$ charges attached, the corresponding expressions are
\begin{align}
    &\mu = \d^ru\wedge\d^r\bar{u} \wedge \d^rD\ \left. \frac{\partial^{2r}}{\partial\lambda^1\dots\partial\lambda^r\partial\gamma^1\dots\partial\gamma^r} Z\ \right|_{\lambda=\gamma=0}\;,\label{mu_formula}\\
    &\mu_{\mathsf{Q}_1,\dots,\mathsf{Q}_r} = \left. \d^ru \wedge \frac{\d \mathsf{Q}_1(D)\wedge \dots\wedge \d \mathsf{Q}_r(D)}{\mathsf{Q}_1(D)\dots \mathsf{Q}_r(D)} Z\ \right|_{\lambda=\gamma=0}\;.
\end{align}
Note that $\mu$ gives the integrand of \eqref{Ell-middle} while $\mu_{\mathsf{Q}_1...\mathsf{Q}_r}$ only contains the 1-loop contributions. The elliptic genus \eqref{Ell-middle} is then
\begin{equation}
    \int_{\mathfrak{C}\times\mathfrak{M} \backslash H^\epsilon} \mu\;,\label{mu-integration}
\end{equation}
where $\mathfrak{C}$ is the integration interval of $D$ to be specified shortly.

As was shown in \cite{Benini:2013nda,Benini:2013xpa}, given a family of hyperplanes $H_i^\epsilon$, one can construct a cell decomposition of $\mathfrak{M}\backslash H^\epsilon$:
\begin{equation}
    \mathfrak{M} \backslash H^\epsilon = \bigsqcup_i C_i\;,\quad\partial C_{i_1\dots i_k} = \sum_jC_{i_1\dots i_k j}- S_{i_1\dots i_k}\;,\label{cell_decomposition}
\end{equation}
where
\begin{equation}
    S_{i}\coloneq \partial H^\epsilon \cap \partial H^\epsilon_i\;,\quad S_{i_1\dots i_s}=S_{i_1}\cap\dots\cap S_{i_s}\;.
\end{equation}
We also need to specify $\mathfrak{C}$. The simplest choice would be $\mathbb{R}^r$, but we can shift $\mathbb{R}^r$ by $\text{i}\delta$ with $\delta\in\mathbb{R}^r$.  This is safe as long as $\delta$ is small enough so that the $D$-contour does not touch any poles in this shift\footnote{Note that there are no poles when $D\in\mathbb{R}^r$ in \eqref{mu_formula}. Therefore, it is safe to shift $\mathfrak{C}$ infinitesimally in \eqref{mu-integration}. However, poles at $D=0$ will be introduced after we apply the Stokes relation and obtain the integrands like $\mu_{Q_1,\dots, Q_p}$. As a result, our choice of $\delta$ would change the selection of the poles, but the total contribution should be kept invariant.}.

In the following we will choose $\mathsf{Q}=Q$,
which guarantees that the differential forms 
$\mu_{\mathsf{Q}_1, \dots, \mathsf{Q}_s}$
are non-singular in $\mathfrak{C}\times \mathfrak{M}\backslash H^{\epsilon}$.
(Recall that $H^{\epsilon}$ is defined from $Q$, not from $\mathsf{Q}$.) 
In this case, by invoking \eqref{iteration_formula} and \eqref{cell_decomposition}, one can use the Stokes relation iteratively to simplify the integral.

As an example, we have
\begin{equation}
   \int_{\mathfrak{C}\times\mathfrak{M} \backslash H^\epsilon} \mu = \sum_i\int_{\mathfrak{C} \times \partial C_i}\d\mu_{Q_i} =-\sum_i\int_{\mathfrak{C}\times S_i}\mu_{Q_i} + \sum_{i\neq j} \int_{\mathfrak{C}\times C_{ij}} \mu_{Q_i}\;.
\end{equation}
The second term can be further simplified to integrals over $S_{ij}$. To understand better how the shift $\delta$ would influence the result, let us pause here and take a closer look at the first term.

\begin{figure}[ht]
    \centering
    \includegraphics[width=15cm]{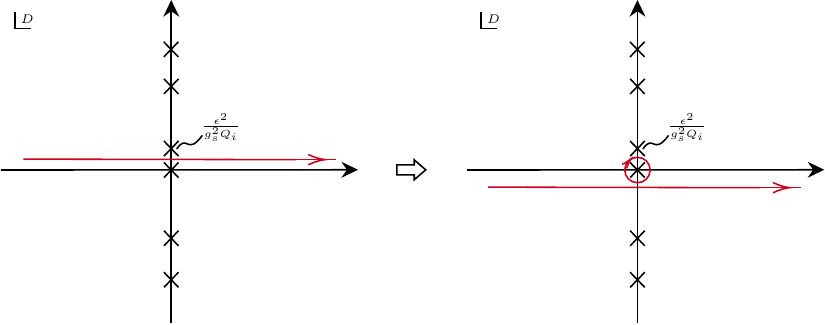}
    \caption{The contour for $D$. In this illustration, there is a pole in the upper half plane approaching the origin when $\epsilon\rightarrow0$. This pushes the contour to the lower half plane with an extra circle surrounding the origin.}\label{r=1_D_contour}
\end{figure}

For simplicity, let us first consider the $r=1$ case. In this case, we only have the terms of form $\int_{\mathfrak{C}\times S_i} \mu_{Q_i}$. This integral can be written as
\begin{equation}
    \oint_{\partial H_i^\epsilon} \d u \int_{\mathbb{R}+\text{i}\delta} \d D  \ f\left(u,\overline{u},D\right) \frac{1}{D}\prod_{m,n}\frac{m+n\overline{\tau} +\rho_i\left(\overline{u},\overline{\xi}\right)}{|m+n\tau+\rho_i(u,\xi)|^2 + \text{i} Q_i D}\;. \label{r=1_integral}
\end{equation}
Here, $f(u,\bar{u},D)$ is regular in $H_i^\epsilon$. For the $D$-integral, the poles are at $D=0$ and $D=\text{i}|m+n\tau+\rho_i(u,\xi)|^2/Q_i$. Around $\partial H_i^\epsilon$, there is one pole at $D=\text{i}\epsilon^2/Q_i$, and the other poles are finitely away from $D=0$ in the limit $\epsilon\to0$. If $Q_i\delta<0$, the $u$-integral would identically give zero as there is no pole in $u$.  However, if $Q_i\delta>0$, the pole at $D=\text{i}\epsilon^2/Q_i$ would hit the contour in the limit $\epsilon\to0$. Therefore, we must shift the contour through $D=0$ as in Figure \ref{r=1_D_contour}, which leaves a circular contour around $D=0$.
Denote the infinite product in \eqref{r=1_integral} as $g\left(u,\overline{u},D\right)$. We have
\begin{equation}
    \eqref{r=1_integral} = 
    \begin{cases}
        \displaystyle\int_{\partial H_i^\epsilon}\d u f\left(u,\overline{u},0\right)g\left(u,\overline{u},0\right) \propto \text{Res}_{u=u_i}\ f(u)g(u)\;, & Q_i\delta>0\;,\\
        0\;, & Q_i\delta<0\;.
    \end{cases}
\end{equation}

This picture can be generalized to higher ranks as long as the vectors $Q_i$ are linearly independent. For each singularity coming from a scalar field with ``charge'' $Q_i$, if $Q_i(\delta)<0$, the result would be zero; if $Q_i(\delta)>0$, we shall deform $\mathfrak{C}$ to $\mathfrak{C}_i\cong S^1\times\mathbb{R}^{r-1}$, where $S^1$ is the contour surrounding $Q_i(D)$. This manipulation can be performed iteratively to get the $D$-contour of form $\mathfrak{C}_{i_1\dots i_l}\simeq T^r$ surrounding $Q_{i_1}(D),\ \dots,\ Q_{i_l}(D)$. 

In general, it is technically complicated to keep track of all the possible contours, integrands, and signs. Nevertheless, the argument from \cite{Benini:2013xpa} still works. Following the prescription, given an arbitrary vector $\eta\in(\mathbb{R}^r)^*$, the elliptic genus reads
\begin{equation}
    I(\tau,\xi) = \frac{1}{|W|}\sum_{i_1,\dots,i_r}c_{i_1\dots i_r,\eta}\int_{S_{i_1\dots i_r} } \d^ru \ Z_{1-\text{loop}}(u,\xi)\;, 
\end{equation}
where 
\begin{equation}
    c_{i_1\dots i_r,\eta}=
\begin{cases}
    1\;, & \eta \in \text{Cone}(Q_{i_1},\dots,Q_{i_r})\;, \\
    0\;, & \text{otherwise}\;.
\end{cases}
\end{equation}
We remark that the result should not depend on $\eta$, since a change of $\eta$ is equivalent to changing the imaginary shift of the $D$-contour. However, there is no pole at $D=0$ before we apply the Stokes relation. Therefore, we can perform any infinitesimal imaginary shift on the $D$-contour without changing the result.

\subsection{Cases with \texorpdfstring{$b\neq r$}{b≠r}}\label{bneqr}
So far, we have assumed $b=r$, i.e., the number of uncharged Fermi multiplets is equal to the rank of the gauge group $G$. Now, let us discuss the situations with $b\neq r$. We shall see that we would obtain zeros for $b<r$, while it would be much more complicated for $b>r$.

Let us begin with $b<r$. In this case, we construct a new function
\begin{equation}
    Z_{i_1\dots i_{r-b}} =\frac{\partial^{r-b}}{\partial \lambda_{i_1}\dots\partial\lambda_{i_{r-b}}} Z\;.
\end{equation}
With \eqref{Stokes}, we can derive the relation
\begin{align}
    \frac{\partial}{\partial\overline{u}^a}Z_{i_1\dots i_{r-b}}&= \frac{\partial^{r-b}}{\partial\lambda_{i_1}\dots\partial\lambda_{i_{r-b}}}\left(\frac{\partial Z}{\partial\overline{u}^a}\right)\nonumber \\
    &=\frac{\partial^{r-b+2}Z}{\partial \lambda_{i_1}\dots\partial\lambda_{i_{r-b}} \partial\lambda_{a}\partial\gamma^l}D^l-\frac{\partial}{\partial\overline{u}^{i_k}}\frac{\partial^{r-b}}{\partial\lambda_{i_1}\dots\widehat{\partial\lambda_{i_{k}}} \partial\lambda_a\partial\lambda_{i_{k+1}}\dots\partial \lambda_{i_{r-b}}}Z\;.\label{ubar-rel}
\end{align}
In parallel to the above discussions, we define an $(r-n)$-form in the $\overline{u}$-space:
\begin{align}
    \Omega'_{a_1\dots a_n} =& \frac{1}{(r-n)!^2} \d \overline{u}^{c_1} \wedge \dots \wedge \d \overline{u}^{c_{b-n}} \wedge\d \bar{u}^{i_1}\wedge \dots \wedge \d\overline{u}^{i_{r-b}} \epsilon_{d_1 \dots d_{b-n} a_1 \dots a_n} \nonumber \\
     &\left.\frac{\partial^{2(b-n)}}{\partial \lambda^{c_1} \partial \gamma^{d_1} \dots \partial \lambda^{c_{b-n}} \partial \gamma^{d_{b-n}}} Z_{i_1\dots i_{r-b}}\ \right|_{\lambda=\gamma=0}\;.\label{generalized-omega-def}
\end{align}
Combining \eqref{generalized-omega-def} with \eqref{ubar-rel}, it is straightforward to see that \eqref{partial-D-relation} holds for $\Omega'$. Hence,  the definition \eqref{mu-Q-def} and the differential relation \eqref{iteration_formula} of $\mu_{\mathsf{Q}_1,...,\mathsf{Q}_s}$ can also be applied here by replacing $\Omega$ with $\Omega'$. Now, we have
\begin{align}
    &\mu' = \d^ru\wedge \d^r\overline{u} \wedge \d^rD \left. \frac{\partial^{2r}}{\partial\lambda^1\dots\partial\lambda^r\partial\gamma^1\dots\partial\gamma^r} Z\ \right|_{\lambda=\gamma=0}\;,\\
    &\mu'_{\mathsf{Q_1},\dots,\mathsf{Q_b}} = \d u^{i_1}\wedge \d\overline{u}^{i_1} \wedge \dots \wedge \d u^{i_{r-b}}\wedge d \overline{u}^{i_{r-b}}\wedge \d^bu \wedge \left.\frac{\d\mathsf{Q}_1(D)\wedge \dots \wedge \d\mathsf{Q}_b(D)}{\mathsf{Q}_1(D)\dots\mathsf{Q}_b(D)}Z_{i_1\dots i_{r-b}}\ \right|_{\gamma=\lambda=0}\;.
\end{align}

The cell decomposition procedure \eqref{cell_decomposition} would still work, and we can use the Stokes relation to iteratively simplify the elliptic genus to the integrals of $\mu'_{Q_1,\dots, Q_b}$ over $(r,r-b,b)$-cycles in the $(u,\overline{u}, D)$-space. However, terms of form $\left.\frac{\partial Z}{\partial\lambda_{i_i}\dots\partial \lambda_{i_k}}\ \right|_{\lambda=\gamma=0}$ would always vanish. This can be seen from the perturbative expansion of
\begin{equation}
    \left\langle\exp\left(\int \d^2z \psi^i\left(z,\overline{z}\right)\lambda_a(T^a)_{ij}\phi^j\left(z,\overline{z}\right)\right)\right \rangle_0\;.
\end{equation}
After performing the weight decomposition on $V^s\cong \bigoplus\limits_{i} V^s_{i}$, this gives
\begin{equation}
    \left\langle\exp\left(\int \d^2z \sum_{k}\psi_k\left(z,\overline{z}\right) \rho_k(\lambda)\phi_k\left(z,\overline{z}\right)\right)\right \rangle_0\;.
\end{equation}
Expanding it in powers of $\lambda$, one can find that all the terms would vanish after Wick contractions. Hence, after applying the Stokes relation, we would obtain a complicated expression of the form
\begin{equation}
    \sum_k \int_{C_k} \mu'_{Q_1,\dots, Q_b} =0
\end{equation}
since the integrands are identically zero.

For the cases with $b>r$, the spirit is similar. We define
\begin{equation}
    Z_{j_1\dots j_{b-r}} = \frac{\partial^{b-r}}{\partial\gamma_{j_1}\dots\partial\gamma_{j_{b-r}}}Z\;.\label{b-r-function}
\end{equation}
Then we can perform the same procedure as above. Instead of duplicating similar expressions, let us give a brief discussion here. In this case, we can use the Stokes relation to simplify the integral to an integral over the $(r,0,r)$-cycles plus $(0,0,b-r)$ flat directions in the $(u,\overline{u}, D)$-space. The integrand in this case would be a $(b-r)$-point function of form \eqref{b-r-function}. However, it is no longer a holomorphic function even when $D=0$. In general, \eqref{b-r-function} is also non-zero, and its concrete expression depends on the concrete form of the $J$-terms. We shall not delve deeper into this problem. It may be interesting to ask if it can still be simplified to some residue-type formula.

\section{The Gukov-Pei-Putrov Model}\label{sec:GPP}
In this section, we illustrate the utility of our formula with the theories studied in \cite{Gukov:2019lzi}. For the SO gauge group cases, they can be summarized in the quiver diagram
\begin{equation}
    \includegraphics[width=7.5cm]{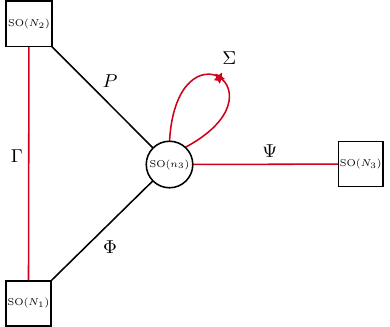}\;,
\end{equation}
where a black (resp.~red) edge stands for a chiral (resp.~Fermi) multiplet. A line connecting two nodes indicates that it transforms as the vector representations under both of the groups, while a loop with a star represents a (reducible) symmetric representation. The superpotential reads
\begin{equation}
    \bigintss\d^2x\,\d\theta\,\left(\sum_{\alpha,\beta=1}^{n_3}\Sigma^{\alpha\beta}\left(A\sum_{i=1}^{N_1}\Phi^{\alpha}_i\Phi^{\beta}_i+B\sum_{l=1}^{N_2}P^{\alpha}_lP^{\beta}_l-C\delta^{\alpha\beta}\right)+\sum_{i=1}^{N_1}\sum_{l=1}^{N_2}\sum_{\alpha=1}^{n_3}\Gamma_{i,l}\Phi^{\alpha}_iP^{\alpha}_l\right)
\end{equation}
for some real constants $A$, $B$ and $C$. The cancellation of the perturbative anomaly requires
\begin{equation}
    n_3=\frac{N_1+N_2-N_3}{2}\;.
\end{equation}

Let us write down the path integral expression for the elliptic genus. We require $n_3$ and $N_{1,2,3}$ to be all even, since otherwise the elliptic genus is zero due to the simple reason that the Pfaffian of an odd-dimensional square matrix is zero. This agrees with the conclusion in \cite{Gukov:2019lzi}. Now, for the gauge and flavour groups all of D-type, we have
{\small\begin{align}
    I=&\frac{1}{2^{n_3/2-1}(n_3/2)!}\sum_{u^*}\text{Res}_{u^*}(\eta)\left(\prod_{\alpha=1}^{n_3/2}\d u_\alpha\right)\left(\frac{2\pi\eta(\tau)^2}{\text{i}}\right)^{n_3/2}\nonumber\\
    &\left(\prod_{\alpha<\beta}^{n_3/2}\frac{\text{i}\,\theta_1\left(s_{1,\alpha\beta}u_\alpha+s_{1,\beta\alpha}u_\beta\right)}{\eta(\tau)}\frac{\text{i}\,\theta_1\left(s'_{1,\alpha\beta}u_{\alpha}+s'_{1,\beta\alpha}u_\beta\right)}{\eta(\tau)}\right)\nonumber\\
    &\left(\prod_{\alpha>\beta}^{n_3/2}\frac{\text{i}\,\theta_1\left(s_{2,\alpha\beta}u_\alpha+s_{2,\alpha\beta}u_\beta\right)}{\eta(\tau)}\frac{\text{i}\,\theta_1\left(s'_{2,\alpha\beta}u_\alpha+s'_{2,\beta\alpha}u_\beta\right)}{\eta(\tau)}\right)\left(\prod_{\alpha=1}^{n_3/2}\frac{\text{i}\,\theta_1\left(2s_{2,\alpha}u_\alpha\right)}{\eta(\tau)}\right)\nonumber\\
    &\left(\prod_{\alpha=1}^{n_3/2}\prod_{a=1}^{N_3/2}\frac{\text{i}\,\theta_1\left(s_{3,\alpha a}u_\alpha+t_{3,a\alpha}\nu_a\right)}{\eta(\tau)}\frac{\text{i}\,\theta_1\left(s'_{3,\alpha a}u_\alpha+t'_{3,a\alpha}\nu_a\right)}{\eta(\tau)}\right)\left(\prod_{i=1}^{N_1/2}\prod_{r=1}^{N_2/2}\frac{\text{i}\,\theta_1\left(\lambda_i-\mu_r\right)}{\eta(\tau)}\frac{\text{i}\,\theta_1\left(\lambda_i+\mu_r\right)}{\eta(\tau)}\right)\nonumber\\
    &\left(\prod_{\alpha=1}^{n_3/2}\prod_{i=1}^{N_1/2}\frac{\text{i}\,\eta(\tau)}{\theta_1\left(s_{4,\alpha i}u_\alpha+t_{1,i\alpha}\lambda_i\right)}\frac{\text{i}\,\eta(\tau)}{\theta_1\left(s'_{4,\alpha i}u_\alpha+t'_{1,i\alpha}\lambda_i\right)}\right)\nonumber\\
    &\left(\prod_{\alpha=1}^{n_3/2}\prod_{r=1}^{N_2/2}\frac{\text{i}\,\eta(\tau)}{\theta_1\left(s_{5,\alpha r}u_\alpha+t_{2,r\alpha}\mu_r\right)}\frac{\text{i}\,\eta(\tau)}{\theta_1\left(s'_{5,\alpha r}u_\alpha+t'_{2,r\alpha}\mu_r\right)}\right)\;,
\end{align}}
where $\lambda_i$, $\mu_r$ and $\nu_a$ are the chemical potentials for the flavour groups $\text{SO}(N_1)$, $\text{SO}(N_2)$ and $\text{SO}(N_3)$ respectively. We choose $\eta=(1,\dots,1)$, and $Q_i$ has $A/C$ (resp.~$B/C$) at the $i^\text{th}$ entry and zero elsewhere for the factors from $\Phi$ (resp.~$P$). Here, $s$, $s'$, $t$ and $t'$ with various subscripts are signs such that the corresponding factors give the Pfaffians and such that the integrand is a well-defined function on the torus satisfying the bi-periodicity condition. Therefore,
\begin{align}
\begin{split}
    &s_{1,\alpha\beta}+s'_{1,\alpha\beta}+s_{2,\alpha\beta}+s'_{2,\alpha\beta} \\
    &\quad =2s_{2,\alpha}+\sum_{a=1}^{N_3/2}\left(s_{3,\alpha a}+s'_{3,\alpha a}\right)-\sum_{i=1}^{N_1/2}\left(s_{4,\alpha i}+s'_{4,\alpha i}\right)-\sum_{r=1}^{N_2/2}\left(s_{5,\alpha r}+s'_{5,\alpha r}\right)=0\;.
\end{split}
\end{align}
For the signs $t$ and $t'$, they would simply put relations among the chemical potentials associated with the flavour symmetries. Nevertheless, the chemical potentials can still be independent by setting $t_*+t'_*=0$ for each pair of $t_*$ and $t'_*$ with the same subscript. A choice of the signs can be
\begin{align}
    I=&\frac{1}{2^{n_3/2-1}(n_3/2)!}\sum_{u^*}\text{Res}_{u^*}(\eta)\left(\prod_{\alpha=1}^{n_3/2}\d u_\alpha\right)\left(\frac{2\pi\eta(\tau)^2}{\text{i}}\right)^{n_3/2}\left(\prod_{\alpha<\beta}^{n_3/2}\frac{\text{i}\,\theta_1\left(u_\alpha+u_\beta\right)}{\eta(\tau)}\frac{\text{i}\,\theta_1\left(u_{\alpha}-u_\beta\right)}{\eta(\tau)}\right)\nonumber\\
    &\left(\prod_{\alpha>\beta}^{n_3/2}\frac{\text{i}\,\theta_1\left(u_\alpha+u_\beta\right)}{\eta(\tau)}\frac{\text{i}\,\theta_1\left(u_\alpha-u_\beta\right)}{\eta(\tau)}\right)\left(\prod_{\alpha=1}^{n_3/2}\frac{\text{i}\,\theta_1\left(2u_\alpha\right)}{\eta(\tau)}\right)\nonumber\\
    &\left(\prod_{\alpha=1}^{n_3/2}\prod_{a=1}^{N_3/2}\frac{\text{i}\,\theta_1\left(u_\alpha+\nu_a\right)}{\eta(\tau)}\frac{\text{i}\,\theta_1\left(u_\alpha-\nu_a\right)}{\eta(\tau)}\right)\left(\prod_{i=1}^{N_1/2}\prod_{r=1}^{N_2/2}\frac{\text{i}\,\theta_1\left(\lambda_i-\mu_r\right)}{\eta(\tau)}\frac{\text{i}\,\theta_1\left(\lambda_i+\mu_r\right)}{\eta(\tau)}\right)\nonumber\\
    &\left(\prod_{\alpha=1}^{n_3/2}\prod_{i=1}^{N_1/2}\frac{\text{i}\,\eta(\tau)}{\theta_1\left(u_\alpha+\lambda_i\right)}\frac{\text{i}\,\eta(\tau)}{\theta_1\left(u_\alpha-\lambda_i\right)}\right)\left(\prod_{\alpha=1}^{n_3/2}\prod_{r=1}^{N_2/2}\frac{\text{i}\,\eta(\tau)}{\theta_1\left(u_\alpha+\mu_r\right)}\frac{\text{i}\,\eta(\tau)}{\theta_1\left(u_\alpha-\mu_r\right)}\right)\;.
\end{align}
In particular, this integrand verifies the conjectured one in \cite{Yagi:2024tac}\footnote{\cite{Yagi:2024tac} discussed a solution to the tetrahedron equation \cite{MR611994}
in the spirit of the Gauge/YBE correspondence \cite{Spiridonov:2010em,Yamazaki:2012cp,Yamazaki:2013nra,Yamazaki:2018xbx}.}. Again, we remark that different compatible choices of signs of the Pfaffians would only differ the elliptic genus by an overall sign convention.

As analyzed in \cite{Gukov:2019lzi}, the theory has different phases that depend on the relative signs of $A$, $B$, and $C$. In particular, from the path integral perspective, it is straightforward to see that the elliptic genus is invariant under the simultaneous change of the signs $(A, B, C)\rightarrow(-A,-B,-C)$ since the poles to be chosen would only be determined by the signs of $C/A$ and $C/B$. We will now see that they can be recovered by taking different residues.

\paragraph{A toy model} As a warm-up, let us first consider a degenerate case where $N_2=N_3=0$. Then $n_3=N_1/2$. When $C/A<0$, we do not take any poles following our prescription. Therefore, the elliptic genus is simply 0, indicating a phase where supersymmetry is broken. When $C/A>0$, our prescription tells us that we should take all possible combinations of $n_3/2$ poles out of $\theta_1(u_\alpha\pm\lambda_i)$. There are $\displaystyle\binom{n_3^2}{n_3/2}$ such choices. Denote $\mathfrak{S}$ as a subset of $\{1,\dots,N_1/2\}$ and $\overline{\mathfrak{S}}$ its complement. Then\footnote{Notice that the number of summands is less than the number of all possible pole choices. This is because some of the combinations have residue zero due to the factors such as $\theta_1(u_\alpha+u_\beta)$ and $\theta_1(u_\alpha-u_\beta)$ in the numerator.}
\begin{equation}
    I=2\sum_{\substack{\mathfrak{S}\subseteq\{1,\dots,N_1/2\}\\|\mathfrak{S}|=n_3/2}}\left(\prod\limits_{i\in\mathfrak{S}}\prod\limits_{j\in\overline{\mathfrak{S}}}\frac{\text{i}\,\eta(\tau)}{\theta_1(\lambda_j-\lambda_i)}\frac{\text{i}\,\eta(\tau)}{\theta_1(\lambda_j+\lambda_i)}\right)=0\;,\label{Itoymodel}
\end{equation}
where the last equality follows from the global residue theorem. Therefore, it would be natural to expect that the phase with $C/A>0$ can be continuously deformed to the one with $C/A < 0$, where supersymmetry is broken.

One may also compute the elliptic genus using the geometric description as in \cite{Gukov:2019lzi}. The NLSM in the IR has target space being the real oriented\footnote{By the real oriented Grassmannian $\widetilde{\text{Gr}}(k,n)$, we mean the manifold consisting of all oriented $k$-dimensional subspaces of $\mathbb{R}^n$, which is a double cover of the real (unoriented) Grassmannian $\text{Gr}(k,n)$.} Grassmannian $\widetilde{\text{Gr}}(n_3,N_1)$ with the trivial bundle of left-moving fermions. As a result, the elliptic genus computed therefrom is exactly \eqref{Itoymodel}.

\paragraph{The phase with \texorpdfstring{$C/A>0$}{C/A>0} and \texorpdfstring{$C/B<0$}{C/B<0}} Let us now consider the cases where all $N_i$ are non-trivial. When $C/A>0$ and $C/B<0$, we should take poles only from the scalar multiplets charged under $\text{SO}(N_1)$ but not $\text{SO}(N_2)$. The elliptic genus is
\begin{align}
    I=&2\sum_{\substack{\mathfrak{S}\subseteq\{1,\dots,N_1/2\}\\|\mathfrak{S}|=n_3/2}}\left(\prod_{i\in\mathfrak{S}}\prod_{a=1}^{N_3/2}\frac{\text{i}\,\theta_1(\lambda_i-\nu_a)}{\eta(\tau)}\frac{\text{i}\,\theta_1(\lambda_i+\nu_a)}{\eta(\tau)}\right)\left(\prod_{j\in\overline{\mathfrak{S}}}\prod_{r=1}^{N_2/2}\frac{\text{i}\,\theta_1(\lambda_j-\mu_r)}{\eta(\tau)}\frac{\text{i}\,\theta_1(\lambda_j+\mu_r)}{\eta(\tau)}\right)\nonumber\\
    &\left(\prod_{i\in\mathfrak{S}}\prod_{j\in\overline{\mathfrak{S}}}\frac{\text{i}\,\eta(\tau)}{\theta_1(\lambda_j-\lambda_i)}\frac{\text{i}\,\eta(\tau)}{\theta_1(\lambda_j+\lambda_i)}\right)\;.
\end{align}
This agrees with the elliptic genus obtained in \cite{Gukov:2019lzi} from the NLSM description where the target space is $\widetilde{\text{Gr}}(n_3, N_1)$ and the bundle of the left-moving fermions is $S^{N_3}\oplus Q^{N_2}$. Here, $S$ and $Q$ denote the tautological and the normal bundles over the real oriented Grassmannian, respectively. In particular, the summand coming from each fixed point would also match.

\paragraph{The phase with \texorpdfstring{$C/A<0$}{C/A<0} and \texorpdfstring{$C/B>0$}{C/B>0}} In this phase, we should take poles only from the chirals charged under $\text{SO}(N_2)$ but not $\text{SO}(N_1)$. It turns out that the elliptic genus has the same expression as in the previous case, with $N_1\leftrightarrow N_2$ and $\lambda\leftrightarrow\mu$ exchanged.

As checked in \cite{Gukov:2019lzi}, the elliptic genera for this phase and the above phase would differ by a sign. It is natural to conjecture that they differ by an orientation reversal such that $I\left(\mathcal{T}\cup\overline{\mathcal{T}}\right)=I(\mathcal{T})+I\left(\overline{\mathcal{T}}\right)=0$. Notice that although there is a sign ambiguity in defining the elliptic genus, once the sign convention for $\mathcal{T}$ is fixed, its orientation reversal $\overline{\mathcal{T}}$ would also have its elliptic genus fixed.

\paragraph{The phase with \texorpdfstring{$C/A<0$}{C/A<0} and \texorpdfstring{$C/B<0$}{C/B<0}} It was argued in \cite{Gukov:2019lzi} that this is the supersymmetry broken phase. Indeed, we do not take any poles in this phase as $C/A<0$ and $C/B<0$. Therefore, the elliptic genus is zero.

\paragraph{The phase with \texorpdfstring{$C/A>0$}{C/A>0} and \texorpdfstring{$C/B>0$}{C/B>0}} We should take the sum of all the Grothendieck residues when $C/A>0$ and $C/B>0$. As a result,
\begin{equation}
    I=0
\end{equation}
following our discussions in \cref{res}. Therefore, we expect that this phase is connected to the supersymmetry broken phase with $C/A<0$ and $C/B<0$.

Let us comment on the computation from the NLSM description in this phase. It was obtained in \cite{Gukov:2019lzi} that the classical target space is 
\begin{align}
    \bigsqcup\limits_{k=0}^{n_3}\frac{\widetilde{\text{Gr}}(k,N_1)\times\widetilde{\text{Gr}}(n_3-k,N_2)}{\mathbb{Z}_2} \;,
\end{align}
where the $\mathbb{Z}_2$ action is the diagonal deck transformation that forgets the orientations of the $l$-dimensional subspaces in $\mathbb{R}^n$ and hence sends each $\widetilde{\text{Gr}}(l,n)$ to $\text{Gr}(l,n)$. For each connected component in the disjoint union, the bundle of the left-moving fermions is 
\begin{align}
    \frac{S_1^{N_3}\oplus S_2^{N_3}\oplus(S_1\otimes S_2)\oplus(Q_1\otimes Q_2)}{\mathbb{Z}_2} \;,
\end{align}
where $S_1$, $Q_1$ and $S_2$, $Q_2$ are the tautological and the normal bundles over $\widetilde{\text{Gr}}(k,N_1)$ and $\widetilde{\text{Gr}}(n_3-k,N_2)$ respectively. From the NLSM, we have\footnote{Notice that $k$ is always even in the sum.}
\begin{align}
    I_\text{cl}=&\sum_{\substack{\mathfrak{S}_1\subseteq\{1,\dots,N_1/2\}\\\mathfrak{S}_2\subseteq\{1,\dots,N_2/2\}\\|\mathfrak{S}_1|=k/2\\|\mathfrak{S}_2|=(n_3-k)/2}}\left(\prod_{i\in\mathfrak{S}_1}\prod_{a=1}^{N_3/2}\frac{\text{i}\,\theta_1(\lambda_i-\nu_a)}{\eta(\tau)}\frac{\text{i}\,\theta_1(\lambda_i+\nu_a)}{\eta(\tau)}\right)\left(\prod_{r\in\mathfrak{S}_2}\prod_{a=1}^{N_3/2}\frac{\text{i}\,\theta_1(\mu_r-\nu_a)}{\eta(\tau)}\frac{\text{i}\,\theta_1(\mu_r+\nu_a)}{\eta(\tau)}\right)\nonumber\\
    &\left(\prod_{i\in\mathfrak{S}_1}\prod_{r\in\mathfrak{S}_2}\frac{\text{i}\,\theta_1(\lambda_i+\mu_r)}{\eta(\tau)}\frac{\text{i}\,\theta(\lambda_i-\mu_r)}{\eta(\tau)}\right)\left(\prod_{j\in\overline{\mathfrak{S}_1}}\prod_{s\in\overline{\mathfrak{S}_2}}\frac{\text{i}\,\theta_1(\lambda_j+\mu_s)}{\eta(\tau)}\frac{\text{i}\,\theta_1(\lambda_j-\mu_s)}{\eta(\tau)}\right)\nonumber\\
    &\left(\prod_{i\in\mathfrak{S}_1}\prod_{j\in\overline{\mathfrak{S}_1}}\frac{\text{i}\,\eta(\tau)}{\theta_1(\lambda_j-\lambda_i)}\frac{\text{i}\,\eta(\tau)}{\theta_1(\lambda_j+\lambda_i)}\right)\left(\prod_{r\in\mathfrak{S}_2}\prod_{s\in\overline{\mathfrak{S}_2}}\frac{\text{i}\,\eta(\tau)}{\theta_1(\mu_s-\mu_r)}\frac{\text{i}\,\eta(\tau)}{\theta_1(\mu_s+\mu_r)}\right)\;,
\end{align}
which is in general not zero.

We propose that there are corrections to this geometric description at the quantum level (cf.~\cref{dynamics})\footnote{Similar discussions can be found, for example, in \cite{Melnikov:2019tpl} for the $(2,2)$ and $(0,2)$ cases.}. While the coupling constants in the superpotential are classically marginal, there could be non-trivial RG-running for $C$ due to quantum corrections\footnote{To the best of our knowledge, we are not aware of any non-renormalization theorem proven in the $(0,1)$ case. Therefore, there could even be corrections at all loops (although it could still be possible that the only divergent diagram contributing to $C$ is the 1-loop tadpole correction).}. We expect the beta function to be proportional to the product of the 1-loop correction to the expectation value of the auxiliary scalar in the symmetric Fermi multiplet coming from the tadpole diagram and the total Cartan charges of the scalars in the chiral multiplets under the gauge group. More concretely, following \eqref{RGFlow}, we have
\begin{equation}
    \mu\frac{\partial}{\partial\mu}C(\mu)=n_3(N_1A+N_2B)\;.
\end{equation}
Moreover, $A$ and $B$ would only receive UV divergent corrections from the diagrams containing the scalar field self-energy loops, which do not change the sign of $A$ and $B$. Therefore, the phase with $C/A>0$ and $C/B>0$ would be connected to the supersymmetry broken phase under the RG flow.

Now, the NLSM computation does not agree with the GLSM and the RG flow computations. This could be explained as follows. So far, we have assumed that the target space is smooth as a disjoint union. If we consider the space $J^{-1}(0)$ before being quotiented by the gauge symmetry, there would be singular loci in this space, as can be seen using the Jacobian criterion. As the $\text{SO}(n_3)$ rotation acts freely, leaving no singular locus fixed, the space should still be singular after modding out the gauge symmetry. Therefore, the actual space could be the gluing of the pieces in the above disjoint union, and singularities could arise at the glued parts where $k$ jumps. Then each smooth part separated by the singular loci would pick up an extra $\pm1$ sign due to the $\mathbb{Z}_2$ line bundle from $\nabla J$. Moreover, there would be contributions from the singular loci, which would make things more complicated. Nevertheless, after taking these into account, we expect that the integration localized to $J^{-1}(0)$ in the strong coupling limit would agree with the weak coupling and the RG flow computations.

It is not surprising that the elliptic genus could have some phase transition phenomena when $A=0$ or $B=0$ (with $N_i$ all non-trivial). As we saw in the path integral derivation, these quadratic terms in the $J$-term provide extra mass terms that regularize the infinite contributions from the bosonic zero modes. However, when $A=0$ or $B=0$, the corresponding quadratic terms would vanish, and the bosonic zero modes would provide some flat directions in the moduli space. As a result, along these loci, there are contributions from infinitely far points in the field configuration space, and the standard argument of the invariance of the elliptic genera from supersymmetric localization would fail.

\begin{figure}[ht]
    \centering
    \includegraphics[width=0.49\linewidth, trim = 60 0 80 0]{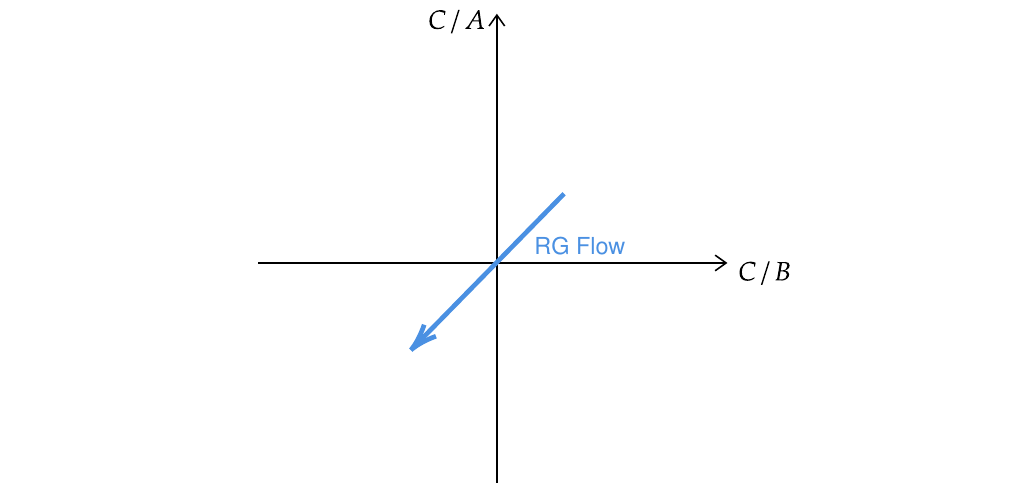}
    \includegraphics[width=0.49\linewidth, trim = 60 0 140 150]{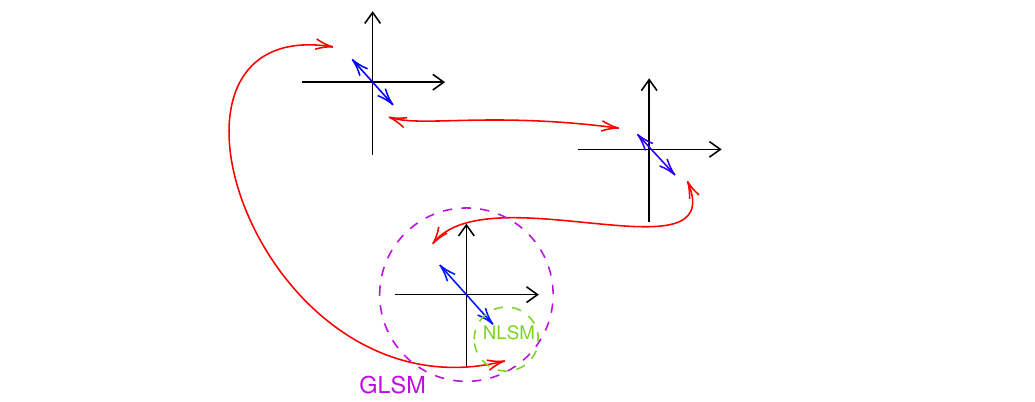}
    \caption{The left picture indicates the flow from the upper right phase to the lower right supersymmetry breaking phase. In the right picture, theories related by trialities are schematically shown with the red arrows. Each pair of upper left and lower right corners connected by a blue arrow has elliptic genera that differ by a sign only, which could be interpreted as different choices of orientations.}\label{phases}
\end{figure}

Let us summarize the phase structure in \cref{phases}. The left picture indicates that the upper right phase can be deformed to the lower left supersymmetry-breaking phase. In the right picture, the red arrows are the dual theories with different ranks of the gauge and flavour groups. This is known as the triality \cite{Gadde:2013lxa}, and the details can be found in \cite{Gukov:2019lzi}. The blue arrows give the phases (with the same gauge and flavour groups) that could be understood as differing by an orientation.

\paragraph{Comparison with the \texorpdfstring{$(0,2)$}{(0,2)} triality} Let us comment on the distinction between our theories and the $(0,2)$ triality in \cite{Gadde:2013lxa}. In the latter model, the $(0,2)$ vector multiplet contains a $(0,1)$ adjoint Fermi multiplet. As stated in \cref{comparisons}, this renders $Q_i$ equal to $\rho_i^G$, the roots of $G$, so that the residues coincide with the JK residues. In the GPP model here, there is no such adjoint Fermi multiplet, and $Q_i$ are determined by the generic quadratic terms in the $J$-terms.

\paragraph{The \texorpdfstring{$\text{O}(n)$}{O(n)} and \texorpdfstring{$\text{USp}(n)$}{USp(n)} gauge groups} The elliptic genus can be computed in the same manner when we change the gauge group. For the $\text{O}(n)$ gauge group, one needs to introduce two extra Fermi multiplets transforming under the determinant representation of the gauge group to cancel the anomaly. This would introduce some extra constant factors to the elliptic genus. For the $\text{USp}(n)$ gauge group, the symmetric Fermi multiplet is replaced by the anti-symmetric Fermi multiplet. The computations are similar to the discussions above. Therefore, we shall omit the explicit expressions here.

\section{Discussions}\label{sec:discussions}
In this paper, we have derived new exact expressions for the elliptic genera of $\mathcal{N}=(0,1)$ GLSMs. We have also applied the formula to the GPP model and discussed the phase structure of the theory.

Since our formula is general, it would be interesting to apply it to various theories, such as those in \cite{Franco:2021ixh,Franco:2021vxq}, and test the duality/triality. Moreover, we plan to report the study of this formula using a more geometric approach. This would allow us to write down the expressions for the theories whose neutral fermionic zero modes are larger than the rank of the gauge group, as well as theories other than GLSMs. It is also a natural question to ask whether the formula for the relevant topological invariants in this paper can be derived via virtual localizations with a rigorous mathematical ground (cf.~\cite{ontani2023virtual}).

The Stolz-Teichner conjecture states that 2d $\mathcal{N}=(0,1)$ theories can be classified by the TMFs up to continuous deformations \cite{segal1987elliptic,stolz2004elliptic,stolz2011supersymmetric}. In particular, the deformation invariants of the theories should be more refined than the (unflavoured) elliptic genera. Since then, there have been extensive studies on this in physics \cite{bunke2009secondary,Gukov:2018iiq,Gaiotto:2019gef,Tachikawa:2021mvw,Tachikawa:2021mby,Lin:2021bcp,Tachikawa:2023nne,Tachikawa:2023lwf,Johnson-Freyd:2024rxr,Tachikawa:2025flw}. We hope that the flavoured/equivariant elliptic genera considered here could shed light on the discussions of the flavoured compact SCFTs and the equivariant TMFs \cite{gepner2023equivariant,Lin:2024qqk}.

Finally, in the study of BPS state counting in $\mathcal{N}\geq (0,2)$ quiver gauge theories, Jeffrey-Kirwan residues have been used \cite{Bao:2024ygr,Bao:2025hfu} (see also \cite{Nekrasov:2017cih,Nekrasov:2018xsb,Nekrasov:2023nai}) in the derivation of the statistical mechanical model of crystal melting \cite{Okounkov:2003sp,Szendroi:2007nu,Ooguri:2009ijd,Ooguri:2009ri,Yamazaki:2010fz,Yamazaki:2022cdg,Galakhov:2023vic,Franco:2023tly}. It would be interesting to explore the counterparts for $\mathcal{N}=(0,1)$ theories.

\section*{Acknowledgements}
We are indebted to Kentaro Hori, Yuan Miao, Shota Saito, Yanming Su, Yuji Tachikawa, Junya Yagi, and Yi Zhang for many fruitful discussions. This research was supported by the World Premier International Research Center Initiative (WPI), MEXT, Japan. JB and MY were supported in part by JSPS Grant-in-Aid for Scientific Research (Grant No.~23K25865). MY was also supported in part by the JSPS Grant-in-Aid for Scientific Research (Grant No.~23K17689), and by JST, Japan (PRESTO Grant No.~JPMJPR225A, Moonshot R\&D Grant No.~JPMJMS2061). DZ was supported by FoPM, WINGS Program, and the University of Tokyo.

\appendix

\section{Real Representations of Real Lie Algebras}\label{realliealg}
For 2d $\mathcal{N}=(0,1)$ supersymmetry, the supercharge is real and all fields are in general real fields. In particular, the representations of the gauge groups do not necessarily have any complex structures. In this appendix, we shall review some useful facts of the real Lie algebras and their real representations. Readers are referred to \cite{onishchik2004lectures,knapp1996lie,Henneaux:2007ej} for more discussions on the real Lie algebras.

In a real representation of a real semisimple Lie algebra, there is in general no weight decomposition with respect to the action of the Cartan subalgebra, due to the fact that $\mathbb{R}$ is not algebraically closed. To classify the irreducible real representations of the real semisimple Lie algebras, we need to consider the notions of complexifications and realifications.

Let $\mathfrak{g}_0$ be a real semisimple Lie algebra. The representation $\rho:\mathfrak{g}_0\rightarrow\mathfrak{gl}(V_0)$ in a real vector space $V_0$ is called a \emph{real representation} of $\mathfrak{g}_0$. One may also replace $V_0$ with a complex vector space $V$, and $\rho:\mathfrak{g}_0\rightarrow\mathfrak{gl}(V)$ will be called a \emph{complex representation} of the real Lie algebra $\mathfrak{g}_0$. If $\mathfrak{g}$ is a complex Lie algebra, then its representation in a complex vector space is complex.

Given a real vector space $V_0$, we may construct its complexification $V=V_0(\mathbb{C})=V_0\otimes_{\mathbb{R}}\mathbb{C}=V_0\oplus\text{i}\,V_0$. Therefore, $\dim_{\mathbb{R}}V_0=\dim_{\mathbb{C}}V$. The complexification of $\mathfrak{g}_0$ is then the complex Lie algebra $\mathfrak{g}=\mathfrak{g}_0(\mathbb{C})$ with the bracket extended as
\begin{equation}
    [x_1+\text{i}\,y_1,x_2+\text{i}\,y_2]=[x_1,y_1]-[y_1,y_2]+\text{i}\,([x_1,y_2]+[y_1,x_2])\;,\quad\forall\ x_1,y_1,x_2,y_2\in\mathfrak{g}_0\;.
\end{equation}
For a real representation $\rho:\mathfrak{g}_0\rightarrow\mathfrak{gl}(V_0)$, there are two \emph{complexifications}. First, we can have a complex representation of $\mathfrak{g}_0$, denoted as $\rho^{\mathbb{C}}:\mathfrak{g}_0\rightarrow\mathfrak{gl}(V)$, by extending $\rho(x)$ to a complex linear operator in $V$ for any $x\in\mathfrak{g}_0$. Second, we can further extend $\rho^{\mathbb{C}}$ to a complex representation of $\mathfrak{g}$, namely $\rho(\mathbb{C}):\mathfrak{g}\rightarrow\mathfrak{gl}(V)$.

Conversely, given a complex vector space $V$, we may regard it as a real vector space $V_{\mathbb{R}}$,
so that we have $\dim_{\mathbb{R}}V_{\mathbb{R}}=2\dim_{\mathbb{C}}V$. Then a complex representation $\rho:\mathfrak{g}_0\rightarrow\mathfrak{gl}(V)$ can be regarded as a representation in the real vector space $V_{\mathbb{R}}$. This is called the \emph{realification}, which gives rise to a real representation $\rho_{\mathbb{R}}:\mathfrak{g}_0\rightarrow\mathfrak{gl}(V_0)$. Moreover, from a complex representation $\rho:\mathfrak{g}\rightarrow\mathfrak{gl}(V)$, we can obtain a real representation $\rho(\mathbb{R}):\mathfrak{g}_{\mathbb{R}}\rightarrow\mathfrak{gl}(V_{\mathbb{R}})$, where $\mathfrak{g}_{\mathbb{R}}$ is the realification of the complex Lie algebra $\mathfrak{g}$.

A real (resp.~complex) representation is said to be irreducible if the representation space does not contain any non-zero proper real (resp.~complex) invariant vector subspace. Two real (resp.~complex) representations are equivalent if there is a real (resp.~complex) isomorphism between the representation spaces. The relations among various real and complex representations are given by
\begin{align}
    &\begin{tabular}{ccccc}
$\rho:\mathfrak{g}_0\rightarrow\mathfrak{gl}(V_0)$ &  & $\rho^{\mathbb{C}}:\mathfrak{g}_0\rightarrow\mathfrak{gl}(V_0(\mathbb{C}))$ &  & $\rho(\mathbb{C}):\mathfrak{g}\rightarrow\mathfrak{gl}(V_0(\mathbb{C}))$ \\ \hline
$\rho$ irreducible & $\Leftarrow$ & $\rho^{\mathbb{C}}$ irreducible & $\Leftrightarrow$ & $\rho(\mathbb{C})$ irreducible \\
$\rho_1\sim\rho_1$ & $\Rightarrow$ & $\rho^{\mathbb{C}}_1\sim\rho^{\mathbb{C}}_2$ & $\Leftrightarrow$ & $\rho_1(\mathbb{C})\sim\rho_2(\mathbb{C})$
\end{tabular}\;,\\
&\begin{tabular}{ccccc}
$\rho_{\mathbb{R}}:\mathfrak{g}_0\rightarrow\mathfrak{gl}(V_{\mathbb{R}})$ &  & $\rho:\mathfrak{g}_0\rightarrow\mathfrak{gl}(V)$ &  & $\rho(\mathbb{C}):\mathfrak{g}\rightarrow\mathfrak{gl}(V)$ \\ \hline
$\rho_{\mathbb{R}}$ irreducible & $\Rightarrow$ & $\rho$ irreducible & $\Leftrightarrow$ & $\rho(\mathbb{C})$ irreducible \\
$(\rho_1)_{\mathbb{R}}\sim(\rho_2)_{\mathbb{R}}$ & $\Leftarrow$ & $\rho_1\sim\rho_2$ & $\Leftrightarrow$ & $\rho_1(\mathbb{C})\sim\rho_2(\mathbb{C})$
\end{tabular}\;.
\end{align}

A real vector space $V_0$ may be endowed with a \emph{complex structure}, which is defined to be an automorphism $\mathcal{J}\in\text{GL}(V_0)$ such that $\mathcal{J}^2=-e$. For instance, a complex vector space $V$ may be viewed as the real vector space $V_{\mathbb{R}}$ with the complex structure $\mathcal{J}:v\mapsto\text{i}\,v$ for any $v\in V_{\mathbb{R}}$. We say that a complex structure $\mathcal{J}$ in a real vector space $V_0$ is \emph{invariant} under a real representation $\rho$ of $\mathfrak{g}_0$ if
\begin{equation}
    \rho(x)\mathcal{J}=\mathcal{J}\rho(x),\quad\forall\ x\in\mathfrak{g}_0\;.
\end{equation}
Then this representation may be regarded as a complex representation of $\mathfrak{g}_0$ in $(V_0,\mathcal{J})$, and the original representation $\rho$ is its realification.

Likewise, we may also define a \emph{real structure} in a complex vector space to be an \emph{anti-involution} $\mathcal{S}: V\rightarrow V$. By an anti-involution we mean that $S$ is
\begin{itemize}
    \item an \emph{anti-automorphism} of $V$, i.e., $\mathcal{S}(cv)=\overline{c}\mathcal{S}v$ for any $c\in\mathbb{C}$, $v\in V$,
    \item and an \emph{involution} of $V$, i.e., $\mathcal{S}^2=e$.
\end{itemize}
A real vector subspace $V_0$ of $V_{\mathbb{R}}$ is called a \emph{real form} of $V$ if $V_{\mathbb{R}}=V_0\oplus\text{i}\,V_0$. In particular, given a complex vector space $V$, its real forms and its real structures are in bijective correspondence. A real form defines a real structure given by $\mathcal{S}(u+\text{i}\,v)=u-\text{i}\,v$ for any $u,v\in V$ which is a \emph{complex conjugation}. A real structure defines a real form given by $V_0=V^S=\{v\in V|Sv=v\}$. A real structure $\mathcal{S}$ in a complex vector space $V$ is said to be \emph{invariant} under a complex representation $\rho:\mathfrak{g}_0\rightarrow V$ if
\begin{equation}
    \rho(x)\mathcal{S}=\mathcal{S}\rho(x),\quad\forall\ x\in\mathfrak{g}_0\;.
\end{equation}
We also say that the corresponding real form $V_0=V^\mathcal{S}$ is invariant, and the real subrepresentation $\rho_0:\mathfrak{g}_0\rightarrow\mathfrak{gl}(V_0)$ of $\rho$ satisfies $\rho_0^{\mathbb{C}}=\rho$.

The irreducible real representations are classified by the following theorem:
\begin{theorem}\label{theorem_real}
    Any irreducible real representation $\rho:\mathfrak{g}_0\rightarrow\mathfrak{gl}(V_0)$ of a semisimple real Lie algebra $\mathfrak{g}_0$ satisfies one of the followings:
    \begin{enumerate}
        \item The representation $\rho$ admits no invariant complex structures. The complex representation $\rho^{\mathbb{C}}$ of $\mathfrak{g}_0$ is irreducible and admits an invariant real structure.
        \item The representation $\rho$ admits an invariant complex structure, and $\rho=\rho'_{\mathbb{R}}$. Here, $\rho'$ is an irreducible complex representation of $\mathfrak{g}_0$ admitting no invariant real structures.
    \end{enumerate}
    Conversely, any real representation $\rho$ satisfying either of the above two conditions is an irreducible representation.
\end{theorem}
When a complex representation $\rho(\mathbb{C})$ admits an invariant real structure, its weights always come in pairs. To see this, notice that
\begin{equation}
    \rho(\mathbb{C})\left(\overline{h}\right)\mathcal{S}v=\mathcal{S}\rho(\mathbb{C})(h)v=\mathcal{S}\lambda(h)v=\overline{\lambda}(h)\mathcal{S}v
\end{equation}
for any $h$ in the Cartan subalgebra of the complex Lie algebra and any vector $v$ in the weight space $V_\lambda$. For a compact real form, we further have $\overline{h}=-h$.

\section{Localization Conditions}\label{localizationconditions}
When we have more than two supercharges, we can always construct the nilpotent supercharges, which are essential to the localization argument. In fact, with a nilpotent supercharge $\mathcal{Q}$, we can write
\begin{equation}
   \partial_\lambda \int \D\Phi\, \text{e}^{-S+\lambda\mathcal{Q}V} =  \int \D\Phi\, \mathcal{Q}V\text{e}^{-S+\lambda\mathcal{Q}V} =  \int \D\Phi\, \mathcal{Q}\left(V\text{e}^{-S+\lambda\mathcal{Q}V}\right) =0\;,
\end{equation}
where we have exploited the condition $\mathcal{Q}S = \mathcal{Q}^2V = 0$. 

However, when we only have one real supercharge, we can only obtain a weaker argument. Now, we have  $\mathcal{Q}_+^2 = \text{i}\,\partial_+$. Hence, to make the localization work, we must have
\begin{equation}
    \mathcal{Q}_+^2V = 0\;.
\end{equation}
Here, $V$ is commonly an integral of the local functional of fields, that is, $V(\Phi) = \int \d^2x\, F(\Phi(x,t))$. Therefore, we should have
\begin{align}
    \begin{split}
        &\int \d^2x\,\text{i}\,(\partial_t+\partial_x)F(\Phi(x,t))\\
        =&\,\text{i}\,\int \d x\, (F(\Phi(x,\beta_1 ))-F(\Phi(x,0))) + \text{i}\,\int \d t\, (F(\Phi(L,t))-F(\Phi(0,t)))\\
        =&\,0
    \end{split}
\end{align}
with respect to the boundary condition in the path integral. The simplest possibility would be to impose the periodic boundary condition on both the time and spatial directions. 

We are mostly interested in the case where the underlying geometry is a torus. This means that the boundary conditions are given by
\begin{align}
    &\Phi(t+\beta,x+L') = \Phi(t,x)\;, \\
    &\Phi(t,x+L) = \Phi(t,x)\;,
\end{align}
where the variables take values in $t\in [0,\beta_1]$ and $x\in [0,L]$. 
It would be heuristic to check whether or not the localization would work for the GLSM. Since $\mathcal{Q}_+^2 = \text{i}\,\partial_+$, it is nilpotent on the functionals with periodic boundary conditions\footnote{More precisely, when we consider a supersymmetric gauge theory, the super vector field $\mathcal{Q}_+$ should be lifted by the superconnection on the worldsheet, and this relation only holds up to a gauge transformation. Nevertheless, in what follows $\delta_{\mathcal{Q}_+}$ only acts on gauge-invariant quantities, so the additional correction is unimportant.}. This is exactly the case for the Witten index and its refined versions, so the localization technique should be applicable. We can try the standard choice given by
\begin{align}
    \begin{split}
        S_\text{loc} &= -g_1 \delta_{\mathcal{Q}_+} \int \d^2x\, \psi  \partial_-\phi - g_2 \delta_{\mathcal{Q}_+} \int \d^2x\, \gamma D -g_3 \delta_{\mathcal{Q}_+} \int\d^2x\, \gamma J \\ 
	&=g_1 \int \d^2x\, \left(-\partial_+\phi \partial_-\phi +  \text{i}\,\psi  \partial_-\psi \right) 
	+ g_2\int \d^2x\, \left( -D^2 + i\gamma\partial_+\gamma  \right) \\ 
     &\quad~ -g_3 \int \d^2x\, \left(DJ-\text{i}\,\gamma\frac{\partial J}{\partial\phi}\psi  \right)\;.
    \end{split}
\end{align}
We find that the whole action is $\mathcal{Q}_+$-exact. Hence, we have the freedom to adjust any coefficients in the action.

Taking the limit $g_1,~g_2 \to \infty$ and $g_3=m$, we find that the bosonic part will get localized around the zero modes $\partial_\mu \phi=0$. Moreover, the contribution from the scalar potential will get suppressed by a $1/g_2$ factor, so we can take $J=0$. Therefore, all the interaction terms can be taken to be 0, and the only object we need to compute is the 1-loop determinants of the fermions and the bosons, integrated over the moduli of the bosonic zero modes. This is exactly the same as in the $(0,2)$ case.

It turns out that the $\mathcal{Q}_+$-exactness of our action can be seen from a simple fact that
\begin{equation}
    \int \d \theta^+\, \ell  
    = \delta_{\mathcal{Q}_+}\ell\ |_{\theta^{+}=0} ,
    \label{exactness_of_action}
\end{equation}
where we have used the fact that $\mathcal{Q}_+|_{\theta^+=0} =  \partial/\partial\theta^+$. Here, $\ell$ should be understood as a superfield constructed out of the polynomials of the ``fundamental'' superfields and their derivatives. Therefore, the ordinary Lagrangian density is always a $\mathcal{Q}_+$-variation of the lowest component of the superfield $\ell$. 

\section{Pfaffians of 2d Chiral Fermions}\label{Pfaffians}
As we are focusing on the case of $(0,1)$ GLSMs, we shall consider the Pfaffians of the 2d chiral fermions coupled to the flat connections (and the RR spin structure). 

Given a 2d spacetime $M$, the moduli space of 2d flat connections is $\text{Hom}(\pi_1(M), G)/G$. Since we are interested in the elliptic genus, we can take $M$ to be a torus with complex structure parameter $\tau$. The moduli space can be further simplified to $\mathbb{C}^{r}/(\mathbb{Z}+\tau\mathbb{Z})^r$, where $r=\text{rank}(G)$. The global anomaly cancellation means that our partition function must be well-defined over this quotient space.

Let us consider the case where we have real chiral fermions $\psi\in S_+(T^2 ) \otimes V$ for some vector space $V$, where $S_+$ denotes the positive chirality spinor bundle. The gauge group $G$ acts on $V$ by representation $\rho$. We choose a Cartan subalgebra $\mathfrak{t}\in \mathfrak{g}$. The action takes the form
\begin{equation}
    S=\int \d^2z \ \psi\left(\overline{\partial} + A\right)\psi\;,
\end{equation}
where $A = u_a T^a$ and $T^a \in \rho(\mathfrak{t})$. One may compute this by
\begin{equation}
    \text{Pf}\left(\overline{\partial}+A\right) \overset{?}{=} \left(\det(\partial+\overline{A})\right)\left(\overline{\partial}+A)\right)^{1/4}\;.
\end{equation}
However, this naive choice would break the holomorphicity of the Pfaffian. From the perspective of the Pfaffian line bundle, we have a fibre bundle over the moduli space with a natural K\"ahler structure. Therefore, the section $\Pfaff\left(\overline{\partial}+A\right)$ should be compatible with it\footnote{The concrete mathematical formulation can be found in \cite{Freed:1986hv}.}. As a result, we first need to take the holomorphic part of $\det\left(\partial+\overline{A}\right)\left(\overline{\partial}+A\right)$ and then take another square root.

More concretely, we take the harmonic function
\begin{equation}
    \phi_{n,m}\left(z,\overline{z}\right) = \exp\left(\frac{\pi}{\text{Im}\tau}\left(n\left(z-\overline{z}\right) + m\left(\tau\overline{z}-\overline{\tau}z\right)\right)\right)\quad(m,n\in\mathbb{Z})
\end{equation}
on $T^2$. Under the weight decomposition, $V$ decomposes into 2-planes with weights $w_i$. The eigenvalues are
\begin{equation}
    \prod_i \prod_{n,m} |n-m\tau\pm w_i(u)|^2\;.
\end{equation}
Hence, the Pfaffian gives
\begin{equation}
    \pm \prod_{i} \prod_{n,m} (n-m\tau+ w_i(u))\;.
\end{equation}
The overall sign here comes from the ambiguity of taking the square root. One may compute this either with the $\zeta$-regularization \cite{DiFrancesco:1997nk} or with the Hamiltonian formalism \cite{Alvarez-Gaume:1986rcs}. The result is
\begin{equation}
   \Pfaff\left(\overline{\partial}+A\right)=\pm\prod_i \frac{\text{i}\,\theta_1(\tau|w_i(u))}{\eta(\tau)}\;.
\end{equation}
For complex fermions, we can replace the Pfaffian with the determinant as long as we 
consider the product over the fermions that are positively charged under the $\text{U}(1)$ phase symmetry. Since the determinants can be worked out explicitly, we can straightforwardly compute the potential global anomaly of a given theory.

\section{Dynamical Aspects of \texorpdfstring{$(0,1)$}{(0,1)} SQFTs}\label{dynamics}
Compared with better-understood 2d $(2,2)$ or $(0,2)$ SQFTs, $(0,1)$ theories are much less understood/constrained. For example, they do not have known non-renormalization theorems or chiral rings due to the lack of supercharges and R-symmetries. Nevertheless, we are still able to discuss some general dynamics.

\paragraph{Supersymmetry breaking} The supersymmetry in 2d $(0,1)$ SQFTs is much less protected as opposed to their higher-dimensional counterparts. The ordinary supersymmetry-breaking condition in the Hamiltonian formalism is
\begin{equation}
    \widehat{\mathcal{Q}}_+\ket{0} \neq 0\;.\label{susy_beaking}
\end{equation}
Since $\mathcal{Q}_+$ is a real supercharge, we have $\widehat{\mathcal{Q}}_+ = \widehat{\mathcal{Q}}_+^\dagger$. Therefore,
\begin{equation}
    \left\langle\psi\left|\,\widehat{\mathcal{Q}}_+^\dagger\widehat{\mathcal{Q}}_+\,\right|\psi\right\rangle=\left\langle\psi\left|\,\widehat{\mathcal{Q}}_+\,\right|\psi\right\rangle = \left\langle\psi\left|\,\left(\widehat{H}-\widehat{P}\right)\,\right|\psi\right\rangle \geq 0
\end{equation}
for any state $|\psi\rangle$. From this, we see that $\left\langle\widehat{H}\right\rangle_0=\left\langle0\left|\,\widehat{H}\,\right|0\right\rangle\geq0$ since $\widehat{P}\,|0\rangle=0$. Hence, the supersymmetry breaking condition \eqref{susy_beaking} is equivalent to $\left\langle0\left|\,\widehat{H}\,\right|0\right\rangle>0$.

Let us illustrate this with some simple examples. The potential term of a $(0,1)$ NLSM comes from the $J$-term in the supersymmetric action. Consider a minimal theory with a Fermi multiplet $\Gamma = \gamma+\theta^+ D$ and a scalar multiplet $\Phi = \phi + \theta^+\psi $, we can write down the action
\begin{equation}
    L_\text{p} = \int \d\theta^+ J(\Phi) \Gamma = J(\phi)D + \gamma\frac{\partial J(\phi)}{\partial \phi}\psi \;.
\end{equation}
After integrating out the auxiliary field $D$, the on-shell Lagrangian gives\footnote{Recall that there is a term $\frac{1}{2}D^2$ from the kinetic term of the Fermi multiplet. Therefore, the equation of motion reads $\partial L_\text{p}/\partial D=D+J(\phi)=0$.},
\begin{equation}
    L_\text{p}^{\text{os}} 
    = - \frac{1}{2}(J(\phi))^2 + \gamma\frac{\partial J(\phi)}{\partial \phi} \psi \;. 
\end{equation}
Hence, the classical potential is given by $(J(\phi))^2$. If $J(\phi)=0$ has a solution, then at the tree level, the system has a supersymmetry-preserving ground state. Otherwise, we say that the supersymmetry is broken at the tree level. As a simplest example, if one sets $J(\phi) = \phi^2-c$, then supersymmetry is broken for $c<0$, but it is preserved for $c\geq0$ at the classical order.

One may also state this result in the off-shell formalism, where the auxiliary field $D$ is solved by an algebraic equation $D = -J(\phi)$. If $J(\phi_c)\neq0$ at the classical vacuum $\phi(x,t)=\phi_c$, then we have
\begin{equation}
    \langle \mathcal{Q}_+\gamma \rangle_0^\text{tree} = \langle D\rangle_0^\text{tree} = -J(\phi_c)\;.
\end{equation}
Therefore, $J(\phi_c)\neq0$ immediately implies that $\langle \mathcal{Q}_+\gamma \rangle_0 \neq 0$, and supersymmetry is spontaneously broken. Beyond the tree order, we have to compute $\langle D\rangle$ to all orders to check if there is supersymmetry breaking. This is a hard task in general. We prefer to look for better indicators for supersymmetry breaking that can be easily computed non-perturbatively. As discussed in this paper, the elliptic genus could be a good candidate for a large family of $(0,1)$ SQFTs.

\paragraph{Renormalization group flows} The behaviour under RG flows is another interesting dynamical aspect of 2d SQFTs. We shall concentrate on the RG flows from the ultraviolet GLSMs here since they are generic enough for constructions of other SQFTs. For $(2,2)$ and $(0,2)$ models, the IR phase structures of GLSMs were investigated extensively in \cite{Witten:1993yc,Distler:1993mk,Hori:2003ic}. They provided illuminating examples of renormalizations and dualities. It is then natural to ask to what extent this story could be extended to $(0,1)$ GLSMs.

Let us consider a generic GLSM. By classical dimensional analysis, the IR limit corresponds to $m\to\infty$. Therefore, the classical IR vacuum should be fixed by the equation $J(\phi)=0$. The fermions would obtain masses from the Yukawa coupling term $\gamma^a\partial J_a/\partial\phi^\mu|_{J^{-1}(0)} \psi ^\mu$. As a result, the IR massless left-moving fermions would correspond to $\ker\left(\partial J /\partial\phi|_{J^{-1}(0)}\right)$. Hence, the IR theory should be identified with the $(0,1)$ NLSM with the target space\footnote{We assume that $\dim(J^{-1}(0)/G)>0$ here. Otherwise, the IR theory may not be an NLSM (similar to the Landau-Ginzburg phase in $(0,2)$ theories). We shall not consider such situations in this paper.}
\begin{align}
    \begin{split}
        \ker\left(\partial J/\partial\phi|_{J^{-1}(0)}\right)\,\rightarrow&\,E\\
        &\downarrow\\
        X=&\, J^{-1}(0)/G
    \end{split}\;.
\end{align}

The analysis above would only hold at the classical level. Now, let us consider how the theory would be modified by the 1-loop corrections under the RG flow \cite{Witten:1993yc}. Let us expand the field $\phi$ as $\phi = \phi_c + \varphi$ around a classical configuration $\phi_c \in J^{-1}(0)$. We have
\begin{equation}
    J^a(\phi) = \left.\frac{\partial J^a}{\partial \phi^\mu}\right|_{\phi_c} \varphi^\mu + \frac{1}{2}
    \left. \frac{\partial^2 J^a}{\partial \phi^\mu \partial \phi^\nu}\right|_{\phi_c} \varphi^\mu\varphi^\nu + \dots\;.
\end{equation}
The first term would provide a mass matrix $M_{\mu\nu} = m^2(\partial J^a/\partial \phi^\mu)(\partial J^a/\partial \phi^\nu)$ for the $\varphi$ fields\footnote{Strictly speaking, such terms come from 4-point vertices of form $\varphi^2D^2$ when we work in the off-shell formalism.} as the potential term reads $(J(\phi))^2$. The second term would give 3-point couplings of form $\varphi \varphi D$. Let us diagonalize $\varphi$ with respect to $M_{\mu\nu}$ with eigenvalues $M_i$. Moreover, we shall denote $\frac{\partial^2J^a}{\partial \phi^\mu \partial \phi^\nu}\bigg|_{\phi_c}$ in this basis as $\mathsf{J}^a_{ij}$.

\begin{figure}[ht]
    \centering
    \includegraphics[width=15cm]{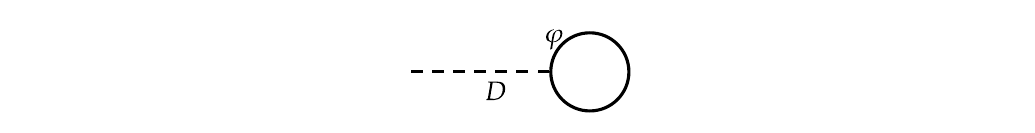}
    \caption{The tadpole diagram for the 1-loop correction of $\langle D \rangle$}.\label{tadpole}
\end{figure}

We would like to compute the 1-loop correction to $-D^a/g_f^2  = mJ^a(\phi)$, i.e., the tadpole diagram of $\delta\langle D^a \rangle$: 
\begin{equation}
    \delta\left\langle-D^a/g_f^2\right\rangle = mC \sum_{i} \mathsf{J}_{ii}^a \int_{0}^{\Lambda} \frac{\d^2p}{p^2+M_i^2} = mC \sum_i\mathsf{J}_{ii}^a\ln(\Lambda/M_i)\;,
\end{equation}
where $C$ is some positive normalization constant and we may set $mC=1$ for simplicity. The tadpole diagram is illustrated in Figure \ref{tadpole}.
Renormalizing it at a scale $\mu$ then gives a running behaviour:
\begin{equation}
    \delta\left\langle -D^a/g_f^2 \right\rangle(\mu) = \sum_i \mathsf{J}_{ii}^a \ln(\mu/M_i)\;.\label{RunningCoupling}
\end{equation}

The correction \eqref{RunningCoupling} does not vanish in general, which implies that either $J^{-1}(0)$ is not the real vacuum specified by $D=0$ of this system, or there is a spontaneous supersymmetry breaking. In the first case, denote the renormalized constants $J^a(0,\mu)$ as $r^a(\mu)$. We have
\begin{equation}
    \mu \frac{\partial} {\partial \mu} r^a(\mu) = \sum_i \mathsf{J}_{ii}^a\;.\label{RGFlow}
\end{equation}
As a simple example, if we consider a model with one scalar and one Fermi multiplet with the quadratic $J$-term $J(\phi) = r + J_1\phi  + J_2 \phi^2$, then classically there always exists a classical vacuum for $r<0$. However, \eqref{RGFlow} implies that $r(\mu)$ would increase in the IR limit when $J_2<0$, and the theory may flow to a supersymmetry-breaking phase\footnote{Notice that the analysis here would only hold at the 1-loop order.}.

\addcontentsline{toc}{section}{References}
\bibliographystyle{ytamsalpha}
\bibliography{references}

\end{document}